\documentclass[10pt,twocolumn]{IEEEtran}
\usepackage[utf8]{inputenc}
\IEEEoverridecommandlockouts 
\usepackage{flushend}
\usepackage{graphicx}
\usepackage{algorithm}
\usepackage{diagbox}
\usepackage{booktabs,amsfonts,dcolumn}
\usepackage{amstext}
\usepackage{tabu,caption}
\usepackage[dvips]{epsfig}
\usepackage{epsfig}
\usepackage[dvips]{graphicx}
\usepackage{multirow}
\usepackage[utf8]{inputenc}
\usepackage[english]{babel}
\usepackage{caption}
\usepackage{subcaption}
\usepackage{float}
\usepackage{ltablex}
\usepackage{cite}
\usepackage{hyperref}
\usepackage{xcolor}
\usepackage{physics}
\usepackage[english]{babel}
\usepackage{cite}
\usepackage{amsmath,amssymb,amsfonts}
\usepackage{algorithmic}
\usepackage{graphicx}
\usepackage{textcomp}
\usepackage{xcolor}
\usepackage{cuted}

\begin{document}

\title{\huge Quantum CDMA-based Continuous Variable Quantum Key Distribution using Chaotic Phase Shifters}

\author{ Shahnoor Ali, Neel Kanth Kundu, and Sourav Chatterjee  \\
\thanks{The work of Neel Kanth Kundu was supported in part by the National Quantum Mission of India, INSPIRE Faculty Fellowship (Reg. No.: IFA22-ENG 344), ANRF Prime Minister Early Career Research Grant (ANRF/ECRG/2024/000324/ENS), and IIT Delhi's New Faculty Seed Grant.}
\thanks{
Shahnoor Ali is with the Center for Applied Research in Electronics (CARE), Indian Institute of Technology Delhi, New Delhi-110016, India (e-mail: shahnoorali1989@gmail.com).

Neel Kanth Kundu is with CARE and the Bharti School of Telecommunication Technology and Management, Indian Institute of Technology Delhi, New Delhi-110016, India (e-mail: neelkanth@iitd.ac.in). He is also an honorary fellow at the Department of Electrical and Electronic Engineering, University of Melbourne, Melbourne, VIC, Australia.

Sourav Chatterjee is a scientist at Tata Consultancy Services -- Research, IIT Kharagpur Research Park, Kolkata-700156, India (e-mail: sourav.chat@tcs.com).
}

}

\maketitle
\begin{abstract}

We present a novel quantum code-division multiple-access (q-CDMA) continuous-variable quantum key distribution (CV-QKD) framework that enables multiuser key distribution over a shared quantum channel. The proposed architecture employs chaotic phase shifters for encoding and decoding quantum information, providing an innovative approach for multiplexing and demultiplexing CV-QKD signals. In the considered approach, quantum states generated by multiple transmitters are chaotically encoded via phase shifters and subsequently multiplexed via a beam splitter (BS) network prior to transmission. At the receiver, a corresponding set of chaotic phase shifters is employed for decoding, followed by demultiplexing via an inverse structure of BSs to recover individual user signals. This process of chaotic synchronization ensures reliable state recovery and secure key establishment between sender--receive pairs. For an arbitrary number of users, we derive the input–output quadrature relations required to characterize the multiuser q-CDMA-based CV-QKD system. Based on this model, we further analyse its achievable secret key rate (SKR) under collective attacks with reverse reconciliation. The influence of critical parameters, including the correction factor, (multiuser) interference noise, environmental noise, and channel transmittance, is systematically investigated. These parameters are shown to play a decisive role in determining their impact on the achievable asymptotic SKR across multiple users. Furthermore, a comparative analysis is provided between the finite-size regime and the asymptotic limit, offering deeper insights into the performance trade-offs. Overall, this study establishes a comprehensive theoretical framework for evaluating the SKR in q-CDMA-based CV-QKD systems, thereby advancing the design of efficient and secure quantum communication networks.

\end{abstract}

\section{Introduction}\label{sec:Introduction}

Quantum key distribution (QKD) is a mature technology that guarantees information‑theoretically secure key exchange between two parties, conventionally referred to as Alice and Bob, based on the fundamental principles of quantum physics rather than computational assumptions\,\cite{gisin2002quantum, pirandola2020advances, xu2020secure}. Unlike classical public‑key cryptosystems relying on the Rivest–Shamir–Adleman (RSA) algorithm\,\cite{rivest1983cryptographic}, which are vulnerable to quantum attacks, the security of QKD systems is grounded in the laws of quantum physics, where any eavesdropping activity inevitably leads to detectable errors\,\cite{gisin2002quantum, xu2020secure}. Furthermore, unlike post‑quantum cryptography\,\cite{liu2024post}--the other leading candidate for mitigating quantum threats--QKD promises perpetual security, thereby eliminating the risk of retroactive vulnerability\,\cite{xu2020secure, papanastasiou2025continuous}. Therefore, QKD will play a major role in ensuring the security and reliability of future communication networks\,\cite{simon2017towards, wei2022towards}.

In a QKD protocol, a symmetric secret key is quantum-mechanically shared between two distant parties -- Alice and Bob, through the exchange of quantum states. Following the transmission and measurement of these quantum states, Alice and Bob communicate over a classical channel to perform post‑processing procedures, including sifting, error correction, and privacy amplification, in order to distill a secure secret key\,\cite{chatterjee2020qkdsim}. To detect the presence of any eavesdropping attack, the classical communication used during post‑processing must be pre-authenticated, ensuring the integrity of the exchanged messages\,\cite{Pirandola:20}.

Quantum key distribution (QKD) was first introduced in 1984 for discrete‑variable (DV) systems\,\cite{bennett2014quantum} and experimentally demonstrated in 1989\,\cite{bennett1992experimental}, wherein information was encoded in discrete degrees of freedom, such as the polarization states of single photons. In DV‑QKD protocols, the extraction of information encoded in polarization\,\cite{bennett1992experimental, chatterjee2020qkdsim}, phase\,\cite{inoue2003differential}, or time‑bin\,\cite{boaron2018secure} degrees of freedom--using single photons or weak coherent pulses--requires detection via single‑photon avalanche detectors, which are technologically demanding and often costly. In contrast, continuous‑variable (CV) QKD protocols were proposed in 2002\,\cite{grosshans2002continuous} as an alternative approach, exploiting continuous degrees of freedom, namely the quadratures of the electromagnetic field. In typical CV‑QKD implementations, information is encoded in the amplitude and phase quadratures of coherent states, commonly modulated according to a two‑dimensional Gaussian distribution\,\cite{jaksch2024composable}. The encoded information is subsequently decoded using homodyne\,\cite{grosshans2002continuous} or heterodyne\,\cite{weedbrook2004quantum} detection, combined with either direct or reverse reconciliation schemes\,\cite{grosshans2002reverse}. Unlike DV‑QKD, these detection techniques rely on off‑the‑shelf components widely employed in classical coherent optical communication systems, rendering CV‑QKD transmitter and receiver architectures largely compatible with existing telecommunication infrastructure\,\cite{pirandola2020advances, zhang2024continuous}. This technological compatibility has been widely recognized as a key factor contributing to the suitability of CV‑QKD for large‑scale deployment in quantum‑secured communication networks\,\cite{lin2019asymptotic}. Prior studies have demonstrated that CV‑QKD offers favorable performance in the asymptotic regime\,\cite{grosshans2003quantum, walenta2014fast}, particularly over short transmission distances\,\cite{papanastasiou2025continuous}. More recent work has further advanced the security analysis\,\cite{leverrier2017security, pirandola2024improved} and experimental realizations of CV‑QKD\,\cite{hajomer2024long, jaksch2024composable, zhang2020long, wang2018high}, leading to substantial improvements in achievable secret key rates and bringing their performance close to the repeaterless Pirandola–Laurenza–Ottaviani–Banchi (PLOB) bound\,\cite{pirandola2017fundamental}. Consequently, contemporary CV‑QKD systems are being increasingly regarded as competitive and cost‑effective alternatives to DV‑QKD implementations for short‑ to metropolitan‑scale distances\,\cite{hajomer2024long, zhang2020long}.

For large‑scale deployment of CV‑QKD in telecommunication networks, efficient multiplexing techniques are required. Prior studies have investigated several approaches for enabling multiuser (i.e., multiple-access) CV‑QKD, including time‑division multiple access (TDMA), frequency‑division multiple access (FDMA), and wavelength‑division multiple access (WDMA). These techniques allocate orthogonal temporal, spectral, or wavelength resources to different users to mitigate inter‑channel interference\,\cite{zhang2024novel, gyongyosi2018multiple, kovalenko2021frequency, zhao2019performance, fang2014multichannel, 10104156}. In contrast, code‑division multiple access (CDMA), which allows multiple users to simultaneously share the entire time and frequency resources through distinct coding sequences, has not yet been thoroughly explored in the context of CV‑QKD protocols. Existing works on CDMA‑based continuous‑variable quantum communication have primarily focused on entanglement distribution\,\cite{q-cdma,rezai2021quantum, PhysRevA.92.042327} rather than secret key generation.


Recently, in our previous work\,\cite{11000234}, we proposed a q-CDMA-based CV-QKD protocol for two users and analyzed the secret key rate (SKR) performance. The results demonstrated that the q-CDMA approach can significantly enhance the SKR performance over long distances compared to conventional CV-QKD schemes without q-CDMA. In this paper, we extend this analysis to a generalized q-CDMA-based CV-QKD network with an arbitrary configuration of $N\times N$ user pairs. We propose a multiplexing and demultiplexing architecture that enables the combination and separation of all quantum signals, respectively, facilitating secure communication between different user pairs. We present a comprehensive analysis of SKR in both asymptotic and finite-size regimes. The main contributions of this article are summarized as follows:  

\begin{itemize}
\item  We propose a novel q-CDMA-based CV-QKD system that supports an arbitrary number of user pairs.
\item We present a multiplexing and demultiplexing scheme based on a binary tree structure of a network of beam splitters (BS). 
\item We derive the input–output relationship between the quadrature of each user pair by incorporating the channel noise and interference noise of the q-CDMA system.
\item We analyse the SKR of the q-CDMA-based CV-QKD system in the asymptotic and finite-size regimes by estimating the channel transmittance and the total excess noise of the quantum channel.  
\item  We rigorously examine the effects of key parameters, such as correction factor, channel length, modulation variance, and environment noise, on the achievable SKR performance -- by conducting extensive numerical simulations.
\end{itemize}

The remainder of the paper is organized as follows. Section~\ref{Sec:2} introduces the system model for the q-CDMA network. Section~\ref{Sec:3} describes the key generation process based on the derived input–output relations for different users, and the asymptotic SKR analysis under homodyne detection, assuming that Eve performs a collective Gaussian attack. In Section~\ref{Sec:4}, we extend the SKR analysis to the finite-size regime, where the channel transmittance and total excess noise are estimated using maximum likelihood estimators. Section~\ref{Sec:5} presents numerical simulation results under experimentally feasible parameter settings, and Section~\ref{Sec:6} concludes the paper with final remarks on the proposed q-CDMA-based CV-QKD system.




{\em Notations}: The creation and annihilation operators for a quantized electromagnetic field are represented by 
$\hat{a}^{\dagger}$ and $\hat{a}$, respectively. Matrices are denoted by boldface uppercase letters ($\boldsymbol{A}$), the transpose of a matrix $\boldsymbol{A}$ is denoted by $\boldsymbol{A}^T$, and the determinant of a matrix is denoted by $\det\boldsymbol{A} $. A Gaussian random variable $X$ with mean $\mu$ and variance $V$ is denoted as $X \sim \mathcal{N}\left(\mu,V \right)$.

\section{System Model for Quantum CDMA based CV-QKD}\label{Sec:2}

\begin{figure*}[t]
    \centering
    \includegraphics[width=0.8\textwidth]{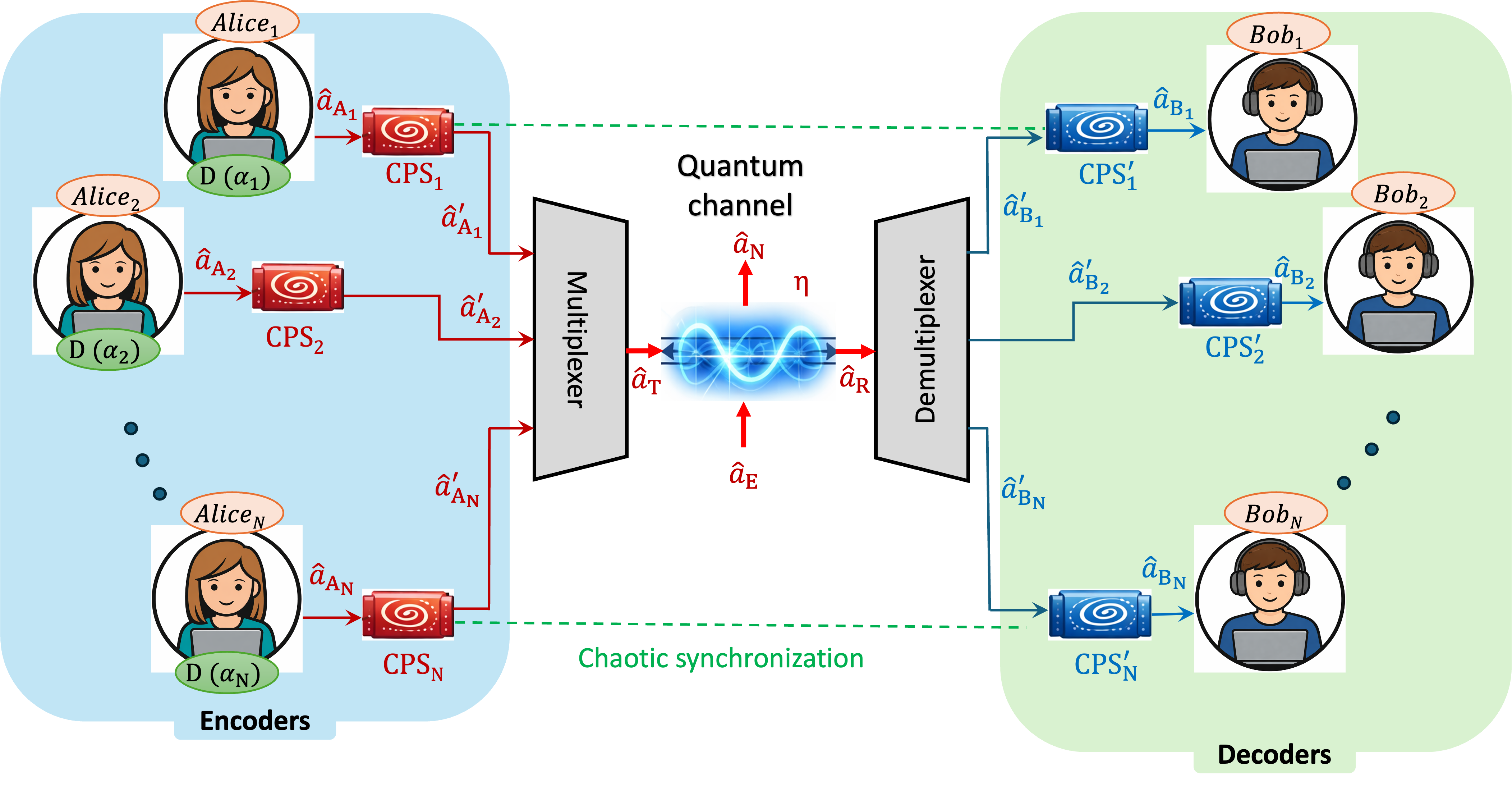}
    \caption{A schematic of a quantum code-division multiple-access (q-CDMA) based continuous variable QKD system using a chaotic phase shifter encoder and decoder.}
    \label{fig:1}
\end{figure*}

Figure~\ref{fig:1} shows a schematic of the proposed q-CDMA-based CV-QKD system.  It comprises $N$ quantum transmitters referred to as $\text{Alice}_{i}$ and $N$ quantum receiver referred to as $\text{Bob}_{i}$ $i\in\{ 1,2,\ldots,N\}$. The quantum signal sent by $\text{Alice}_{i}$ is first encoded by a set of chaotic phase shifters $\text{CPS}_{i}$, and the output signals are combined by a binary tree structure of $N-1$ identical $50:50$ beam splitters and then transmitted via a common quantum channel. The quantum channel is modeled as a beam splitter with transmissivity $\eta$, which introduces loss and potential eavesdropping by an eavesdropper (Eve). At the receiver side, the combined received quantum signal is divided into $N$ branches by another inverse binary tree structure of $ N-1$ identical $50:50$ beam splitters, and sent to $N$ chaotic phase shifters $\text{CPS}^{'}_{i}$, which are introduced to decode the information. The de-multiplexed quantum signals are then sent to the receivers $\text{Bob}_{i}$, which perform homodyne measurement to extract the key information.

The encoding and decoding process of our system could be understood as follows: let us assume that the optical field mode entering into the $i^{th}$ chaotic phase shifter is represented by the annihilation operator $\hat{a}_{A_{i}}$ $(i=1, \ldots, N)$. The operation of the chaotic phase shifters $\text{CPS}_{i}$ can be modeled by the effective Hamiltonian $\hbar\delta_{i}(t)\hat{a}^{\dagger}_{A_{i}}\hat{a}_{A_{i}}$, where $\hbar$ is the Planck constant and $\delta_{i}(t)$ represents frequency of the classical chaotic signal. It has been shown that $\text{CPS}_{i}$ induces a phase shift factor $\exp[-j\theta_{i}(t)]$ in the optical field \cite{q-cdma}. This encoding spreads the quantum signal's spectral content across the entire spectrum. At the receiver's side, the chaotic phase shifters $\text{CPS}^{'}_{i}$ are modeled by the Hamiltonian $-\hbar \delta^{'}(t)\hat{a}'^{\dagger}_{B_{i}}\hat{a}'_{B_{i}}$,
where $\hat{a}'_{B_{i}}$ 
is the annihilation operator of the field mode entering into the $\text{CPS}^{'}_{i}$. This introduces an inverse phase shift factor $\exp[j\theta^{'}_{i}(t)]$ to decode the information-bearing signals masked by the phase shift, where $\theta^{'}_{i}(t) = \int^{t}_{0}\delta^{'}_{i}(\tau)d\tau$. To ensure that the chaotic phase shift applied at the transmitter $\text{Alice}_{i}$ is exactly compensated for at the receiver $\text{Bob}_{i}$, an auxiliary classical channel is established between them to synchronize the two sets of chaotic phase shifters. 

\begin{figure}[htbp]
\centering
\includegraphics[width=0.4\textwidth]{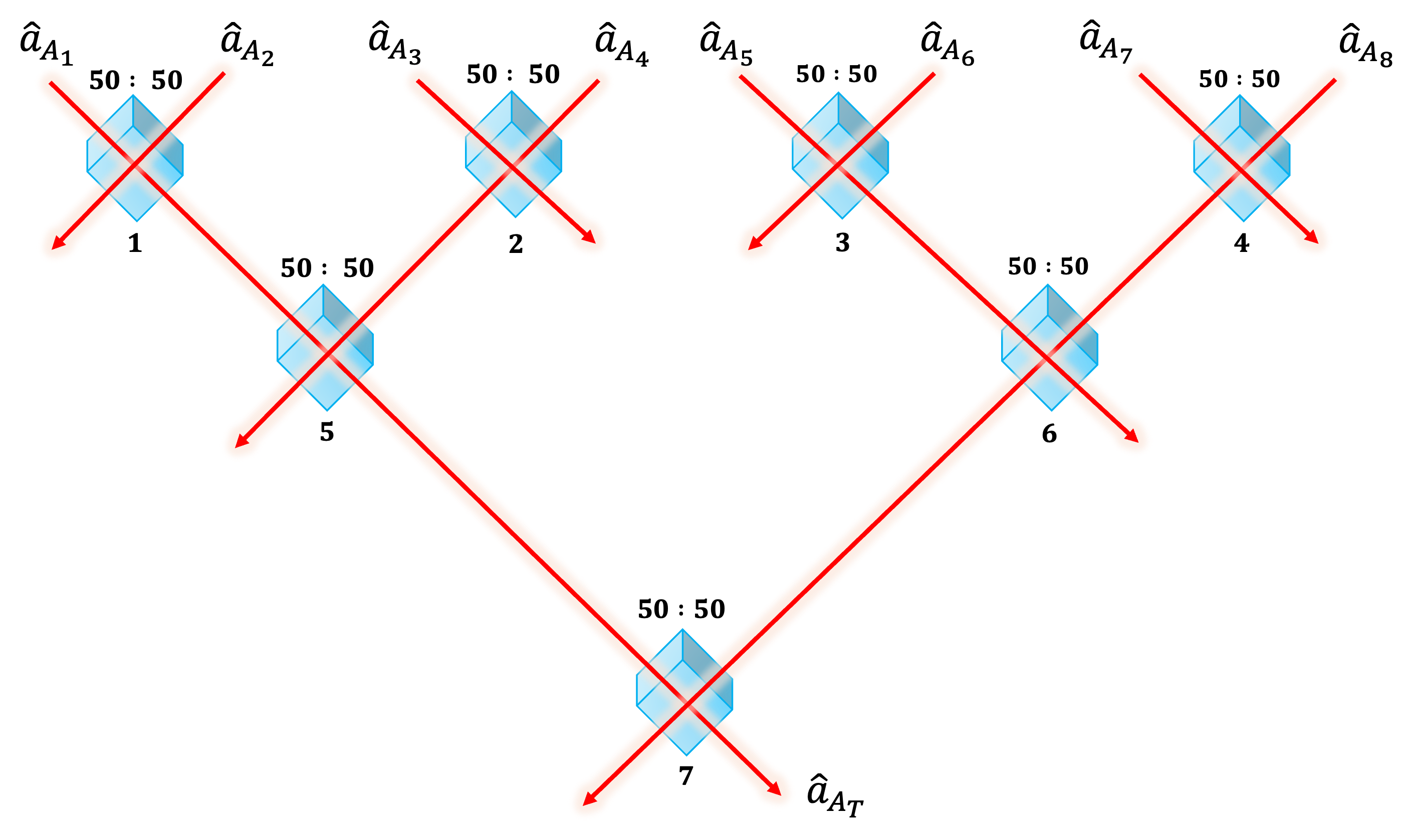}
\caption{Schematic diagram of the multiplexer  stage for $N=8$ transmitters users using seven BSs.}
\label{Fig:Mul}
\end{figure}

\begin{figure}[htbp]
\centering
\includegraphics[width=0.4\textwidth]{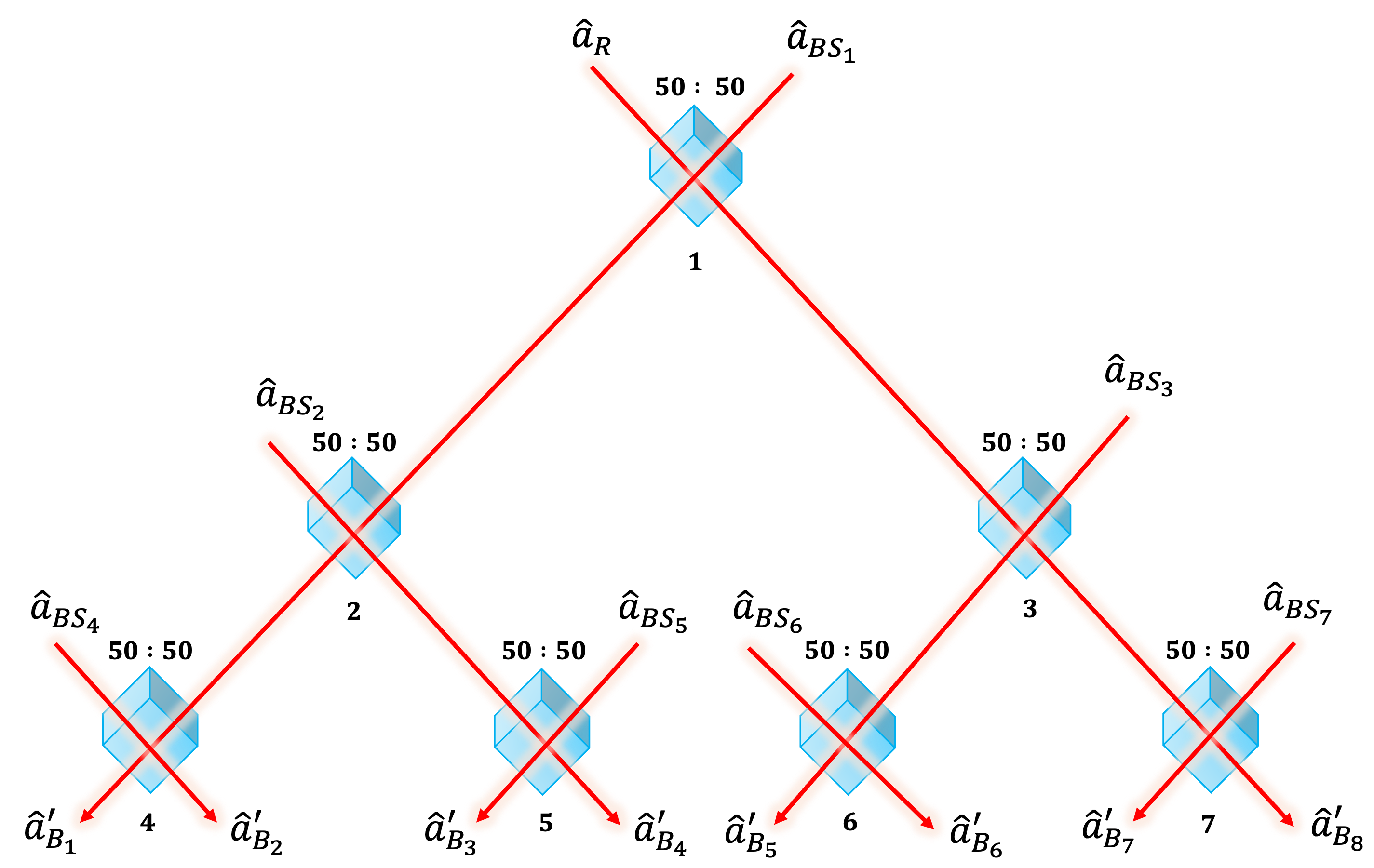}
\caption{Schematic diagram of the de-multiplexing stage for $N=8$ receivers using seven BSs.}
\label{Fig:Dem}
\end{figure}

We first present the binary tree structure-based network of BSs for multiplexing and de-multiplexing for $N=8$ users. Figure.~\ref{Fig:Mul} shows the multiplexing scheme, where a binary tree structure consisting of 7 identical $50:50$ beam splitters is employed to combine all quantum signals emitted by $8$ independent senders $\text{Alice}_{i}$. The resultant output mode of the multiplexer, $\hat{a}_{T}$ of the field mode, represents a coherent superposition of all individual input modes of the signals $\hat{a}^{'}_{A_{1}}, \hat{a}^{'}_{A_{2}}, \ldots, \hat{a}^{'}_{A_{8}}$, expressed as 
\begin{equation}\label{Eq:1}
 \hat{a}_{T} = \sqrt{\frac{1}{8}} \sum_{i=1}^{8} \hat{a}^{'}_{A_{i}}\;. 
\end{equation}
Similarly, Figure~\ref {Fig:Dem} shows the demultiplexing operation, which employs an inverse binary tree structure, comprising seven identical $50:50$ BS's to separate the received composite quantum signal mode $\hat{a}_{R}$. The demultiplexing of the received mode $\hat{a}_{R}$ into the distinct modes  $\hat{a}'_{B_{1}}, \hat{a}'_{B_{2}},  \ldots, \hat{a}'_{B_{8}}$ can be  expressed as
\begin{align}
\hat{a}'_{B_{1}} &= 
 \frac{1}{\sqrt{8}}\,\hat{a}_{R}
+ \frac{1}{\sqrt{8}}\,\hat{a}_{BS_{1}}
+ \frac{1}{2}\,\hat{a}_{BS_{2}}
+ \frac{1}{\sqrt{2}}\,\hat{a}_{BS_{4}}, \nonumber \\[6pt]
\hat{a}'_{B_{2}} &= 
  \frac{1}{\sqrt{8}}\,\hat{a}_{R}
+ \frac{1}{\sqrt{8}}\,\hat{a}_{BS_{1}}
+ \frac{1}{2}\,\hat{a}_{BS_{2}}
- \frac{1}{\sqrt{2}}\,\hat{a}_{BS_{4}}, \nonumber \\[6pt]
\hat{a}'_{B_{3}} &= 
  \frac{1}{\sqrt{8}}\,\hat{a}_{R}
+ \frac{1}{\sqrt{8}}\,\hat{a}_{BS_{1}}
- \frac{1}{2}\,\hat{a}_{BS_{2}}
+ \frac{1}{\sqrt{2}}\,\hat{a}_{BS_{5}}, \nonumber \\[6pt]
\hat{a}'_{B_{4}} &= 
 \frac{1}{\sqrt{8}}\,\hat{a}_{R}
+ \frac{1}{\sqrt{8}}\,\hat{a}_{BS_{1}}
- \frac{1}{2}\,\hat{a}_{BS_{2}}
- \frac{1}{\sqrt{2}}\,\hat{a}_{BS_{5}}, \nonumber \\[6pt]
\hat{a}'_{B_{5}} &= 
 \frac{1}{\sqrt{8}}\,\hat{a}_{R}
- \frac{1}{\sqrt{8}}\,\hat{a}_{BS_{1}}
+ \frac{1}{2}\,\hat{a}_{BS_{3}}
+ \frac{1}{\sqrt{2}}\,\hat{a}_{BS_{6}}, \nonumber \\[6pt]
\hat{a}'_{B_{6}} &= 
 \frac{1}{\sqrt{8}}\,\hat{a}_{R}
- \frac{1}{\sqrt{8}}\,\hat{a}_{BS_{1}}
+ \frac{1}{2}\,\hat{a}_{BS_{3}}
- \frac{1}{\sqrt{2}}\,\hat{a}_{BS_{6}}, \nonumber \\[6pt]
\hat{a}'_{B_{7}} &= 
 \frac{1}{\sqrt{8}}\,\hat{a}_{R}
- \frac{1}{\sqrt{8}}\,\hat{a}_{BS_{1}}
- \frac{1}{2}\,\hat{a}_{BS_{3}}
+ \frac{1}{\sqrt{2}}\,\hat{a}_{BS_{7}}, \nonumber \\[6pt]
\hat{a}'_{B_{8}} &= 
 \frac{1}{\sqrt{8}}\,\hat{a}_{R}
- \frac{1}{\sqrt{8}}\,\hat{a}_{BS_{1}}
- \frac{1}{2}\,\hat{a}_{BS_{3}}
- \frac{1}{\sqrt{2}}\,\hat{a}_{BS_{7}} .
\nonumber
\end{align}
where 
$\{\hat{a}_{BS_{i}}\}_{i=1}^{7}$ 
represent the annihilation operators corresponding to the vacuum fields entering into the unused input port of each BS. These vacuum modes ensure the preservation of commutation relations across all the output modes.

The proposed multiplexing-demultiplexing architecture can be generalized to accommodate an arbitrary $N$ number of user pairs by using $N-1$ identical $50:50$ beam splitters arranged in a binary tree structure. Due to the topology of this structure, the number of users is restricted to a power of two, i.e., $N=2^{q}$, where $q\in \mathbb{Z}$ represents the number of beam splitter levels in the binary tree. 
 
The overall multiplexed output optical mode $\hat{a}_{A_{T}}$ at the transmitter side can thus be expressed as the normalized superposition of all $N$ input modes 
\begin{equation}
\hat{a}_{T} = \sqrt{\frac{1}{N}} \sum_{i=1}^{N}\hat{a}^{'}_{A_i}  \;.
\end{equation}
At the receiver $\text{Bob}_{i}$, the corresponding demultiplexing operation reconstructs the individual modes by successively separating the received composite signal of mode $\hat{a}_{R}$ through the binary tree network of BS. The output annihilation operator of the $i^{\mathrm{th}}$ receiver's ($\text{Bob}_{i}$) mode can be written as
\begin{equation}\label{Eq:demux}
\hat{a}^{'}_{B_{i}}
= \frac{1}{\sqrt{N}}\,\hat{a}_{R}
+ \sum_{\ell=1}^{q} \frac{(-1)^{\,p_\ell(i)}}{\sqrt{2^{\,q-\ell+1}}}\; \hat{a}_{BS_{2^{\ell-1}+m_\ell(i)}}, 
\end{equation} 
where the level index $\ell$ runs from the root node $(\ell = 1)$ to the leaf-splitting (terminal) level $(\ell = q)$. The function $m_{\ell}(i)$ identifies the beam splitter index within a given level, whereas the sign bit function $p_{\ell}(i)$ determines the relative phase of the path (left or right) that is followed by $i^{\mathrm{th}}$ mode in the binary tree. These functions can be defined as
\begin{align}
m_\ell(i) &= \left\lfloor\frac{i-1}{2^{\,q-\ell}}\right\rfloor,~~~~(\ell=1,2,3 \ldots, q)\\   \
p_\ell(i)&= \left( \left\lfloor \frac{i - 1}{2^{\,q - \ell}} \right\rfloor \bmod 2^{\,q - \ell + 1} \right), \\[6pt]
&= 
\begin{cases} 
0, & \text{if the path at level } k \text{ goes left}, \\[4pt]
1, & \text{if the path at level } k \text{ goes right.}
\end{cases}
\end{align}
Accordingly, the BS indices for a binary tree with $q$ hierarchical levels are assigned as
\begin{equation}
BS_{2^{\ell-1}+m_{\ell}(i)}~~\text{for}~~\ell = 1,2,...,q, m_{\ell}(i) = 0,1,2,....,2^{\ell-1} -1
\end{equation}
The resulting beam splitter topology, following a breadth-first indexing scheme, can be represented as
\[
\begin{aligned}
\ell &= 1: && BS_{1} \\
\ell &= 2: && BS_{2},\, BS_{3} \\
\ell &= 3: && BS_{4},\, BS_{5},\, BS_{6},\, BS_{7} \\
&\vdots \\
\ell &= q: && BS_{2^{q-1}}, \dots, BS_{2^{q}-1}
\end{aligned}
\]
Since the vacuum modes associated with the beam splitters are introduced only to preserve the canonical commutation relations among the output modes, their individual contributions are physically indistinguishable. Therefore, for simplicity, we assume that all vacuum inputs entering the unused ports of the beam splitters are identical, i.e., 
\begin{equation}
\hat{a}_{BS_{2^{\ell-1}+m_\ell(i)}} \equiv \hat{a}_{BS}, 
\qquad \forall\, m_{\ell}(i).
\end{equation}
Under this assumption, eq.~(\ref{Eq:demux}) simplifies to 
\begin{equation}\label{eq:demux_simplified}
\hat{a}^{'}_{B_{i}}
= \sqrt{\frac{1}{N}}\,\hat{a}_{\mathrm{R}}
+ \sum_{\ell=1}^{q} 
  \frac{(-1)^{\,p_\ell(i)}}{\sqrt{2^{\,q-\ell+1}}}\
  \hat{a}_{BS}
\end{equation}

The quantum channel is assumed to be fully controlled by the eavesdropper (Eve) and is modeled by a beam-splitter. In this model, Eve replaces the channel loss with a beam splitter of transmittance $\eta$, which represents the effective channel transmissivity. The corresponding input–output relations of the channel are given by
\begin{equation}\label{Eq:10a}
\begin{aligned}
\hat{a}_{R} &= \sqrt{\eta}\,\hat{a}_{T} + \sqrt{1-\eta}\,\hat{a}_{E}, \\
\hat{a}_{N} &= \sqrt{\eta}\,\hat{a}_{E} + \sqrt{1-\eta}\,\hat{a}_{T} .
\end{aligned}
\end{equation}

After the demultiplexing operation, the overall input–output relation of the annihilation operator for the $i$th user pair can be expressed as
\begin{equation}\label{Eq:8}
\begin{aligned}
\hat{a}^{\prime}_{B_i} &= \sqrt{\frac{\eta}{N^{2}}} \sum_{i=1}^{N} \hat{a}^{\prime}_{A_i}
+ \sqrt{\frac{1 - \eta}{N}}\, \hat{a}_{E}
+ \sum_{\ell=1}^{q}\frac{(-1)^{p_\ell(i)}}{\sqrt{2^{\,q-\ell+1}}}\hat{a}_{BS} ,
\end{aligned}
\end{equation}
where $\hat{a}_{E}$ denotes Eve’s ancilla mode injected into the unused input port of the channel beam splitter.
Under the synchronization condition of the chaotic phase shifters $\text{CPS}_{i}$ and $\text{CPS}'_{i}$, the phase fluctuations must satisfy $
\delta_i(t) = \delta'_i(t)$,
which implies identical instantaneous phase modulations, i.e.,
$
\theta_i(t) = \theta'_i(t)$
Accordingly, the field operators after phase modulation are related as
\[
\hat{a}'_{A_i} = \hat{a}_{A_i} e^{-j\theta_i(t)}, 
\qquad
\hat{a}_{B_i} = \hat{a}'_{B_i} e^{j\theta_i(t)} .
\]
Under this synchronization condition, the overall input–output relation for the $i^{\mathrm{th}}$ user pair is given by
\begin{equation}\label{Eq:10}
\begin{aligned}
    \hat{a}_{B_{i}} &= \sqrt{\frac{\eta}{N^2}}\, \hat{a}_{A_{i}} 
    + \sqrt{\frac{\eta}{N^2}} \sum_{\substack{k=1 \\ k \neq i}}^{N} 
      \hat{a}_{A_{k}} e^{j(\theta_i - \theta_k)} \\ 
    &\quad + \sqrt{\frac{1 - \eta}{N}}\, \hat{a}_{E} e^{j\theta_{i}}  
    + \sum_{\ell=1}^{q}\frac{(-1)^{p_\ell(i)}}{\sqrt{2^{\,q-\ell+1}}} 
      \hat{a}_{BS} e^{j\theta_{i}}
\end{aligned}
\end{equation}
The chaotic phase shifters introduce pseudo-noise chaotic phase shift $\theta_{i}(t)$. Taking the average of these broadband random signals gives  
$\overline{\exp(\pm j\theta_{i}(t))} = \sqrt{M_{i}}$, where
$M_{i}$ is the correction factor, given by \cite{zhou2009quantum,PhysRevA.75.063414,bergli2009decoherence}
\begin{align}\label{Eq:12}
M_{i} = \exp \left[ -\pi \int_{\omega_{l_{i}}}^{\omega_{u_{i}}} d\omega \frac{S_{\delta_{i}}(\omega)}{\omega^{2}} \right]
\end{align}
where, $S_{\delta_{i}}(\omega)$ denotes the power spectrum density of the signals of frequencies $\delta_{i}(t)$ and $\omega_{l_{i}}$ and $\omega_{u_{i}}$ are the lower and upper bounds of the frequency band, respectively. The integral accounts for the spectral properties of chaotic phase shifts, leading to the correction factor $M_{i}$ \cite{q-cdma, PhysRevA.92.042327}. Consequently, Eq.~(\ref{Eq:10}) can be simplified as
\begin{equation}\label{Eq:13C}
\begin{aligned}
    \hat{a}_{B_{i}} &= \sqrt{\tfrac{\eta}{N^2}}\,\hat{a}_{A_{i}}
    + \sqrt{\tfrac{\eta}{N^2}} \sum_{\substack{k=1 \\ k \neq i}}^{N} \sqrt{M_{i}M_{k}}\,\hat{a}_{A_{k}} \\
    &\quad + \sqrt{\tfrac{(1-\eta)M_{i}}{N}}\,\hat{a}_{E}
    + \sum_{\ell=1}^{q}\frac{(-1)^{p_\ell(i)}}{\sqrt{2^{\,q-\ell+1}}}\sqrt{M_{i}}\,\hat{a}_{BS}
\end{aligned}
\end{equation}

Considering the broadband frequency spectrum of the chaotic signal, the correction factors $M_{i}$ may become extremely small, which leads to $\hat{a}_{B_{i}} \approx \sqrt{\tfrac{\eta}{N^2}}\,\hat{a}_{A_{i}}$, which implies that the quantum signals transmitted from $\text{Alice}_{i}$ to $\text{Bob}_{i}$ can be completely decoupled from each other even though they are simultaneously transmitted via a common quantum channel. In a q-CDMA–based CV-QKD system, the information-carrying optical fields are spectrally broadened by the chaotic phase shifters, making them unintelligible on the quantum channel. The signals can be correctly recovered at the receiver only by undoing these chaotic phase shifts—i.e., by achieving chaotic synchronization to re-sharpen the quantum signal. This idea is quite similar to the classical CDMA communication systems. 

\section{Asymptotic Secret Key rate Analysis} \label{Sec:3}

\subsection{Key Generation}

We consider the Gaussian-modulated CV-QKD protocol, where $\text{Alice}_{i}$ randomly modulates a vacuum state to create a displaced coherent state $\ket{\alpha_{A_{i}}}$. Here, $\alpha_{A_{i}}$ denotes the complex amplitude of the optical field. This random modulation $\alpha_{A_{i}}=X_{A_{i}}+jY_{A_{i}} $ contains two independent variables $X_{A_{i}}$ and $Y_{A_{i}}$ chosen from two dimension Gaussian distribution with variance $V_{S_{i}}$ and zero mean \cite{grosshans2002continuous}. It is these continuous variables that will ultimately be used to construct a secret key between the user pairs. Subsequently, $\text{Alice}_{i}$ transmits the ensemble of randomly modulated pure coherent states to the corresponding $\text{Bob}_{i}$ through a quantum channel that may be intercepted or manipulated by the eavesdropper, Eve.

At the receiving end, $\text{Bob}_{i}$ makes quadrature measurements on the received signal modes to extract the encoded key information. The measurement can be done either by homodyne detection, which measures a single quadrature component, or heterodyne detection, which measures both quadratures. Here, we assume that $\text{Bob}_{i}$ employs homodyne detection to extract real-valued measurement outcomes of one of the quadrature components. Using Eq.~(\ref{Eq:13C}), the quadrature component measured by $\text{Bob}_{i}$ is given by
\begin{equation}\label{Eq:14}
\begin{aligned}
    X_{B_{i}} &= \sqrt{\frac{\eta}{N^2}}\,X_{A_{i}}
    + \sqrt{\frac{\eta}{N^2}}\sum_{\substack{k=1\\k\neq i}}^{N}\sqrt{M_{i}M_{k}}\,X_{A_{k}} \\
    &\quad+ \sqrt{\frac{(1-\eta)M_{i}}{N}}\,X_{E}
    +  \sum_{\ell=1}^{q}\frac{(-1)^{p_\ell(i)}}{\sqrt{2^{\,q-\ell+1}}}\sqrt{M_{i}}\, X_{BS},
\end{aligned}
\end{equation}
where $X_{A_{i}}$, represents the transmitted quadrature component of $\text{Alice}_{i}$ and $X_{A_{k}}$ represents the quadrature component associated with the interference mode from $\text{Alice}_{k}$, and $X_{E}$ is the Gaussian noise quadrature introduced by the eavesdropper into the quantum channel. Additionally, $X_{BS}$ represents the environmental mode of the beam splitter introduced by the de-multiplexer. In prior studies, this environmental noise was not explicitly modeled and was typically assumed to be negligible (i.e., less than unity) \cite{wang2015detection, PhysRevA.92.042327,q-cdma}. However, in this work, we have explicitly incorporated and analyzed the environmental noise using the binary tree architecture of the BS network.

During the key generation phase, the modes prepared by $\text{Alice}_{i}$ can be described in Heisenberg picture as $X_{A_{i}}= X_{S_{i}}+X_{0}$ where $X_{S_{i}}$, $X_{0}$ denote the quadratures of the modulated signal and thermal mode, respectively. Therefore, the overall variance for each transmitted quadrature component of $i^{th}$ $\text{Alice}_{{i}}$ is given by $V(X_{A_{i}}) = V_{A_{i}} = V_{S_{i}} + V_{0}$, where $V_{S_{i}}$ represents the modulation power and $V_{0}$ is the variance of pure vacuum mode of the initial state of  $\text{Alice}_{i}$. 
The Gaussian noise introduced by Eve has a variance of $V(X_{E})=W$. In contrast, the interference quadratures have the same variances as $\text{Alice}_{i}$ and the variance of the environment quadrature $V(X_{BS})=\sigma$.


\subsection{Eve Attack Mode: Collective Attack}
Analyzing the secret key rates of the q-CDMA network-based CV-QKD protocol involves considering various potential attacks. The most important attack is the collective Gaussian attack. This type of attack is the most powerful eavesdropping strategy, followed by quantum mechanics. Our analysis mainly focuses on the collective Gaussian attack, which is modeled using the entangling cloner attack. In the entangling cloner attack, Eve perfectly replaces the quantum channel between $\text{Alice}_{i}$ and  $\text{Bob}_{i}$ with her quantum channel. She then prepares ancilla modes, which are two-mode squeezed states (Einstein-Podolsky-Rosen (EPR) pair), with variance $W$. The loss of the quantum channel is simulated by a beamsplitter of transmission $\eta$. The EPR beam modes are represented by the operators $\hat{a}_{E}$ and $\hat{a}'_{E}$ in the network. Eve retains one mode of the EPR pair $\hat{a}^{'}_{E}$ in a quantum memory and injects the other mode, $\hat{a}_{E}$, into the quantum channel, resulting in the output mode $\hat{a}_{R}$ which is accessible to the receiver and the output of the other port of the BS mode $\hat{a}_{N}$ is accessible to Eve as shown in Fig.\ \ref{fig:1}. Eve then collectively measures all modes $\hat{a}_{N}$ and $\hat{a}'_{E}$ to extract the maximum information about the key. 

The secret key rate $R_{i}$ for the $i^{\mathrm{th}}$ user pair under reverse reconciliation is given by \cite{grosshans2002reverse}
\begin{equation}\label{Eq:15a}
R_i = I\!\left(X_{A_i}:X_{B_i}\right)
      - \chi\!\left(X_{B_i}:E\right),
\end{equation}
where $I(X_{A_{i}}:X_{B_{i}})$ is the Shannon mutual information between $\text{Alice}_{i}$ and $\text{Bob}_{i}$, and $\chi\!\left(X_{B_i}:E\right)$ is the quantum Holevo information between $\text{Bob}_{i}$'s measurements and Eve's quantum state. The total SKR $R$ of the q-CDMA-based CV-QKD system is given by
\begin{equation}\label{Eq:16}
R = \sum_{i=1}^{N} R_i .
\end{equation}
In (\ref{Eq:15a}), $I(X_{A_{i}}:X_{B_{i}})$ is given by
\begin{align}\label{Eq:16a}
I(X_{A_{i}}:X_{B_{i}}):&= H(X_{B_{i}}) -  H(X_{B_{i}}|X_{A_{i}})
\end{align} 
where 
\begin{align}\label{Eq:17}
H(X_{B_{i}})&=\frac{1}{2}\log_{2}V(X_{B_{i}}),
\end{align}
is the classical Shannon entropy and 
\begin{align} \label{Eq:18}
H(X_{B_{i}}|X_{A_{i}})&=\frac{1}{2}\log_{2}V(X_{B_{i}}|X_{A_{i}})
\end{align}
is the conditional Shannon entropy of $\text{Bob}_{i}$ measurement outcomes \cite{Shannon}. The conditional variance $V(X_{B_{i}}|X_{A_{i}}) $ is given by:
\begin{eqnarray}\label{Eq:19}
V(X_{B_{i}}|X_{A_{i}}) =  V(X_{B_{i}}) -\frac{|\expval{X_{B_{i}}X_{s}}|^{2}}{V_{s}} 
\end{eqnarray}
Using Eq~(\ref{Eq:14}) we can calculate the variance of the quadrature measured by $\text{Bob}_{i}$ and the conditional variance of $\text{Bob}_{i}$'s quadrature given $\text{Alice}_{i}$'s quadrature as
\begin{align}\label{Eq:20a}
V(X_{B_{i}}) &= \frac{\eta}{N^2}V_{A_{i}} 
+ \frac{\eta}{N^2} \sum_{\substack{k=1 \\ k \neq i}}^{N} M_{i} M_{k} V_{A_{k}} \nonumber \\
&\quad + \frac{(1-\eta)}{N} M_{i} W 
+ M_{i} \left( \sum_{\ell=1}^{q}\frac{(-1)^{p_\ell(i)}}{\sqrt{2^{\,q-\ell+1}}} \right)^{2}\sigma,
\end{align}
\begin{align}\label{Eq:20b}
V(X_{B_{i}}|X_{A_{i}}) &= \frac{\eta}{N^2}V_{0A_{i}} 
+ \frac{\eta}{N^2} \sum_{\substack{k=1 \\ k \neq i}}^{N} M_{i} M_{k} V_{A_{k}} \nonumber \\
&\quad + \frac{(1-\eta)}{N} M_{i} W 
+ M_{i} \left( \sum_{\ell=1}^{q}\frac{(-1)^{p_\ell(i)}}{\sqrt{2^{\,q-\ell+1}}} \right)^{2}\sigma.
\end{align}
Thus, the Shannon mutual information is given by
\begin{align}\label{Eq:21}
I(X_{A_{i}}:X_{B_{i}})=\frac{1}{2} \log_{2}\frac{V(X_{B_{i}})}{V(X_{B_{i}}|X_{A_{i}})}
\end{align}
where $V(X_{B_{i}})$ and $V(X_{B_{i}}|X_{A_{i}})$ are given by Eq.~(\ref{Eq:20a}) and Eq.~(\ref{Eq:20b}), respectively.

The Holevo information $\chi(X_{B_{i}}: E)$ which quantifies the maximum mutual information between $\text{Bob}_{i}$'s measurement outcomes and Eve's state quantum state is given by
\begin{equation}
\begin{aligned}\label{Eq:22}
\chi(X_{B_{i}}:E) := S(E)- S(E|X_{B_{i}}) \;,
\end{aligned}
\end{equation}
where $S(E):=S(\rho_{E})$ denotes the von Neumann entropy of Eve's quantum state, and 
$S(E \mid X_{B_{i}}) = S\!\bigl(\rho_{E \mid X_{B_{i}}}\bigr)$
 represents von Neumann entropy of Eve's state conditioned on $\text{Bob}_{i}$'s measurement outcome. 
 
 The von Neumann entropy of a Gaussian state $\rho$ containing $\mathrm{2}$ modes can be written in terms of its symplectic eigenvalues as follows \cite{holevo1999capacity}
\begin{equation}\label{Eq:S}
S(\rho) = \sum^{2}_{\lambda=1}g(\nu_{\lambda})
\end{equation} 
where
\begin{equation}\label{Eq:g}
g(x) := \Big(\frac{x+1}{2}\Big) {\rm log}_2 \Big(\frac{x+1}{2}\Big) - \Big(\frac{x-1}{2}\Big) {\rm log}_2 \Big(\frac{x-1}{2}\Big). 
\end{equation}
The symplectic eigenvalues $\nu_{\lambda}$ of covariance matrix (CM) $\boldsymbol{\Sigma}$ are equal to the standard eigenvalues of $|i\boldsymbol{\Omega}\boldsymbol{\Sigma}|$, where $\boldsymbol{\Omega}$ is the symplectic matrix defined as \cite{RevModPhys.84.621}
\begin{equation}\label{Eq:23}
    \boldsymbol{\Omega}: 
    = \bigoplus_{k=1}^{2} 
    \mqty( 0 & 1 \\ -1 & 0 )
\end{equation}
Thus the Holevo information between $\text{Bob}_{i}$ and Eve in Eq.~(\ref{Eq:22}) can be obtained by calculating the symplectic eigenvalues of their corresponding CMs, which are $\boldsymbol{\Sigma}_{E}$ and $\boldsymbol{\Sigma}_{E|X_{B_{i}}}$, respectively. Eve's CM consisting of two ancilla modes $\hat{a}_{E}$ and $\hat{a}^{'}_{E}$ is given by
\begin{equation} \label{Eq:24}
\boldsymbol{\Sigma}_{E} = \mqty(
E_{V} \boldsymbol{I}_2 & \Phi \boldsymbol{Z} \\
\Phi \boldsymbol{Z} & W \boldsymbol{I}_2)
\end{equation}
where $E_{V} = (1 - \eta)V_{T} + \eta W \;, \Phi = \sqrt{\eta (W^{2} - 1)}$, $V_{T} =\frac{1}{N}\sum^{N}_{i=1}M_{i}V_{i}$ is the average input variance over all users, entering into the quantum channel, $W$ is the variance of Eve's ancilla modes
and 
\begin{equation}\label{Eq:26}
\boldsymbol{I}_{2} = \mqty(
1 & 0 \\
 0 & 1
), ~~\boldsymbol{Z} = \mqty(
1 & 0 \\
 0 & -1) \;.
\end{equation}
Then, Eve's symplectic eigenvalues admit 
\begin{equation} \label{Eq:28}
\nu_{1,2} = \frac{1}{2}[\sqrt{(E_{V} + W)^{2} - 4\eta(W^{2} -1)}\pm (E_{V} - W) ] 
\end{equation}
Thus, the von Neumann entropy of Eve's can be calculated by using Eq.~(\ref{Eq:S}) and Eq.~(\ref{Eq:28}).

The von Neumann entropy of Eve's state, conditioned on $\text{Bob}_{i}$'s measurement outcome $S(E|X_{B_{i}})$, can be obtained from the symplectic eigenvalues of Eve's conditional CM after homodyne measurement, given by \cite{laudenbach2018continuous}
\begin{equation}\label{Eq:31}
\boldsymbol{\Sigma}_{E|X_{B_{i}}} = \boldsymbol{\Sigma}_{E} - (V(X_{B_{i}}))^{-1} \boldsymbol{C} \boldsymbol{\Pi} \boldsymbol{C}^{T} 
\end{equation}
where $\boldsymbol{\Sigma}_{E}$ is defined in Eq.~(\ref{Eq:24}), $V(X_{B_{i}})$ is the variance of the quadrature measured by $\text{Bob}_{i}$ given in
Eq.~(\ref{Eq:20a}), and
\begin{equation} \label{Eq:32}
\boldsymbol{\Pi} :=\mqty(
1 & 0 \\
0 & 0) \;.
\end{equation}

Furthermore, $\boldsymbol{C}$ in Eq.~(\ref{Eq:31}) is a $4 \times 2$ matrix describing the quantum correlation between Eve's modes $\{ X_{N}, X^{'}_{E}\}$ 
and Bob's output mode $X_{B_{i}}$ given by
\begin{equation}\label{Eq:33}
\boldsymbol{C}= \mqty(
\langle X_{N}X_{B_{i}} \rangle \boldsymbol{I}_2 \\
\langle X^{'}_{E}X_{B_{i}} \rangle \boldsymbol{Z}\\)
=\mqty(\Xi_{i} \boldsymbol{I}_2\\
\Psi \boldsymbol{Z}) 
\end{equation}
and the output mode of the quantum channel accessible to Eve is given by Eq.~(\ref{Eq:10a}) as
\begin{equation}
X_{N} = \sqrt{\frac{1-\eta}{N}}\sum^{N}_{i=1}\sqrt{M_{i}}X_{A_{i}} - \sqrt{\eta}X_{E} \;.
\end{equation}
Thus, Eve's conditional covariance matrix $\Sigma_{E|X_{B_{i}}}$ admits
\begin{equation} \label{Eq:35}
\boldsymbol{\Sigma}_{E|X_{B_{i}}} = \mqty(\boldsymbol{A}&\boldsymbol{D}\\\boldsymbol{D}^{T}&\boldsymbol{B})
\end{equation}
where
\begin{subequations}
\begin{align}\label{Eq:36}
\boldsymbol{A} &= \mqty(E_{V}-\frac{\Xi^{2}_{i}}{\beta_{i}} & 0 \\ 0 & E_{V}),\\
\boldsymbol{B} &= \mqty(W-\frac{\Psi^{2}}{\beta_{i}} & 0 \\ 0 & W), \\
\boldsymbol{D} &= \mqty(\Phi-\frac{\Xi_{i}\Psi}{\beta_{i}} & 0 \\ 0 & -\Phi).
\end{align}
\end{subequations}
The entries of the matrices are given by
\begin{align}
\Xi_{i}
&= \sqrt{\frac{\eta(1-\eta)}{N}\,M_{i}}\; W 
- \sqrt{\frac{\eta(1-\eta)M_{i}}{N^{3}}}\; V_{A_{i}} \\ \nonumber
&- \sqrt{\frac{\eta(1-\eta)M_{i}}{N^{3}}}\; \sum_{\substack{k=1 \\ k \neq i}}^{N} M_{k}\,V_{A_{k}}, \\[6pt]
\Psi
&= \sqrt{\frac{(1 - \eta)M_{i}}{N}}\; \sqrt{W^{2} - 1}, \\[6pt]
\beta_{i}
&= \frac{\eta}{N^2}V_{A_{i}} 
+ \frac{\eta}{N^2} \sum_{\substack{k=1 \\ k \neq i}}^{N} M_{i} M_{k}\, V_{A_{k}}
+ \frac{(1-\eta)}{N} M_{i} W \nonumber\\
&\quad + M_{i} \left( \sum_{r=1}^{q} \frac{(-1)^{p_r(i)}}{\sqrt{2^{\,q-r+1}}} \right)^{2}\sigma .
\end{align}
The corresponding symplectic eigen values of $\boldsymbol{\Sigma_{E|X_{B_{i}}}}$ admit
\begin{equation}
\nu_{3,4} = \sqrt{\frac{\Delta \pm \sqrt{\Delta^{2}-4\det\boldsymbol{\Sigma}_{E|X_{B_{i}}}}}{2}},
\end{equation}
where,
\begin{align}
\Delta &= \det\boldsymbol{A} + \det\boldsymbol{B} + 2\det\boldsymbol{D}, \\[6pt]
\det\boldsymbol{A}
&= \frac{E_{V}\big(E_{V}\beta_{i}-\Xi_{i}^{2}\big)}{\beta_{i}}, \\[6pt]
\det\boldsymbol{B}
&= \frac{W\big(W\beta_{i}-\Psi^{2}\big)}{\beta_{i}}, \\[6pt]
\det\boldsymbol{D}
&= \frac{\Phi\big(\Psi\Xi_{i}-\Phi\beta_{i}\big)}{\beta_{i}}, \\[6pt]
\det\boldsymbol{\Sigma}_{E|X_{B_{i}}}
&= \frac{W E_{V} - \Phi^2}{\beta_{i}}\Big( W E_{V} \beta_{i} - W \Xi_{i}^{2} - E_{V} \Psi^{2}
\\ \nonumber&- \Phi^2 \beta_{i} + 2 \Phi \Xi_{i} \Psi \Big),
\end{align}
Thus, the von Neumann entropy of Eve's conditional state is given by
\begin{equation}\label{Eq:39}
S(E|X_{B_{i}}) = \sum_{\lambda=3}^{4} g(\nu_{\lambda}),
\end{equation}
where $g(x)$ is the function defined in Eq.~(\ref{Eq:g}).  
Finally, using Eq.~(\ref{Eq:15a}), Eq.~(\ref{Eq:21}) and Eq.~(\ref{Eq:22}), the total secret key rate $R$ can be computed for various system parameters.

\section{Finite-Size Secret Key Rate Analysis}\label{Sec:4}

Considering finite-size effects~\cite{PhysRevLett.100.200501,PhysRevA.81.062343}, the SKR under reverse reconciliation for the $i^\text{th}$ user is given by
\begin{equation} \label{Eq:40}
R_{i,~\mathrm{finite}} = \frac{n}{K} \Big( \beta I(X_{A_{i}}:X_{B_{i}}) - S_{\epsilon_\mathrm{PE}}(X_{B_{i}} : E) - \Delta(n) \Big), 
\end{equation}
where $\beta$ is the reconciliation efficiency parameter, and $K$ is the total number of signals after quantum transmission and measurement, $n$ is the number of signals used for key distillation, $m = K - n$ is the number of signals allocated for parameter estimation, and $I(X_{A_{i}}:X_{B_{i}})$ is the mutual information between $\text{Alice}_i$ and $\text{Bob}_i$, given by eq.~(\ref{Eq:21}). Furthermore, $S_{\epsilon_\mathrm{PE}}(X_{B_{i}}: E)$ denotes the $\epsilon_\mathrm{PE}$-bounded Holevo information between $\text{Bob}_i$'s measurement outcomes and Eve, and $\Delta(n)$ is the finite-size correction term that accounts for statistical fluctuations and the security requirements of privacy amplification given by ~\cite{9507495,PhysRevA.86.032309,PhysRevA.96.042334, Nitin}:
\begin{equation}\label{Eq:41}
\Delta(n) = \big( 2\,\mathrm{dim}\,\mathcal{H}_X + 3 \big) \sqrt{\frac{\log_2(2/\bar{\epsilon})}{n}} + \frac{2}{n} \log_2 \frac{1}{\epsilon_\mathrm{PA}},
\end{equation}
where $\mathrm{dim}\,\mathcal{H}_X = 2$ denotes the Hilbert space dimension associated with the $X$ quadratures of the raw key, $\bar{\epsilon}$ is the smoothing parameter for the smooth min-entropy, and $\epsilon_\mathrm{PA}$ represents the failure probability of the privacy amplification procedure. The first term in Eq.~(\ref{Eq:41}) characterizes the convergence of the smooth min-entropy under the assumption of independent and identically distributed (i.i.d.) signals, whereas the second term accounts for the residual failure probability of privacy amplification.


\subsection{Parameter Estimation}
We now incorporate the parameter estimation (PE) stage, which is essential for characterizing the quantum channel's transmissivity and excess noise. In this process, it is assumed that
$\text{Alice}_{i}$ and $\text{Bob}_{i}$ jointly and randomly select a subset of $m$ signals from the total of $K$ signals exchanged during the quantum communication session. These $m$ signal pairs, denoted by $(X_{A_{i}l, X_{B_{i}l}})_{l=1}^{m}$, are publicly disclosed and used for estimating the channel parameters in the context of a homodyne detection protocol.

Under the standard assumption of a collective Gaussian attack, the disclosed data are modeled as independent and identically distributed (i.i.d.) samples drawn from a joint Gaussian distribution. The statistical relationship between the variables measured by Alice and Bob can be described by the following linear model \cite{PhysRevA.81.062343,PhysRevResearch.6.023321,PhysRevResearch.3.043014}
 \begin{equation}\label{Eq:42}
 X_{B_{i}} = \sqrt{\eta_{A_{i}}} X_{A_{i}} + z_{A_{i}}
 \end{equation}
where $X_{A_{i}}$ represents Alice's quadrature measurement, $X_{B_{i}}$ represents Bob's corresponding measurement, and $z_{A_{i}}$ is a Gaussian noise variable with zero mean and unknown variance $\sigma_{A_{i}}^2$. This noise term captures the combined effects of channel loss and excess noise introduced during transmission. The effective channel transmissivity experienced by the $i^{\mathrm{th}}$ sub-channel is denoted by $\eta_{A_{i}}$. 

\subsubsection{Estimation of the Transmissivity}

To estimate channel transmissivity, Alice and Bob publicly disclose a subset of their correlated data and construct an estimator $\hat{\eta}_{A_{i}}$ of the true transmissivity $\eta_{A_{i}}$. Based on the maximum likelihood principle, the estimator is obtained from the empirical covariance between their classical variables and is given by \cite{PhysRevResearch.6.023321}

\begin{equation} \label{Eq:PE1}
\hat{\eta}_{A_{i}} =
\left(
\frac{\widehat{C}_{X_{A_{i}} X_{B_{i}}}}
{V_{A_{i}}}
\right)^{2},
\end{equation}
where $\widehat{C}_{X_{A_{i}}X_{B_{i}}}$ is the empirical covariance between Alice's modulation variable $X_{A_{i}}$ and Bob's measurement result $X_{B_{i}}$ given by
\begin{equation}\label{Eq:PE2}
\widehat{C}_{X_{A_{i}}X_{B_{i}}} := \frac{1}{m}\sum^{m}_{L=1} X_{A_{i}L}X_{B_{i}L}
\end{equation}
and $V_{A_{i}}=\mathrm{Var}(X_{A_{i}})$ represents Alice's modulation variance. For analytical convenience, Eq.~(\ref{Eq:PE1}) can be rewritten as
\begin{equation}\label{Eq:PE3}
\hat{\eta}_{A_{i}}
=
\frac{V_{\mathrm{cov}}}{V_{A_{i}}^{2}}
\left(
\frac{\widehat{C}_{X_{A_{i}}X_{B_{i}}}}
{\sqrt{V_{\mathrm{cov}}}}
\right)^{2},
\end{equation}
where $V_{\mathrm{cov}}=\mathrm{Var}(\widehat{C}_{X_{A_{i}}X_{B_{i}}})$ denotes the variance of the covariance estimator. Under the assumption of Gaussian modulation and independent samples, the normalized variable
\(
\widehat{C}_{X_{A_{i}}X_{B_{i}}}/\sqrt{V_{\mathrm{cov}}}
\)
follows a normal distribution with mean
\(
C_{X_{A_{i}}X_{B_{i}}}/\sqrt{V_{\mathrm{cov}}}.
\)
Consequently, its square follows a non-central chi-squared distribution with one degree of freedom and non-centrality parameter
$k_{\mathrm{cn}}=\frac{C_{X_{A_{i}}X_{B_{i}}}^{2}}{V_{\mathrm{cov}}}$. As a result, the estimator $\hat{\eta}_{A_{i}}$ follows a scaled non-central chi-squared distribution. Using the statistical properties of this distribution and retaining the leading-order terms in the large-sample limit, the variance of the transmissivity estimator is obtained as

\begin{equation} \label{Eq:PE4}
\mathrm{Var}(\hat{\eta}_{A_{i}})
=
\frac{4 \eta_{A_{i}}^{2}}{m}
\left(
2 +
\frac{\sigma_{A_{i}}^{2}}
{\eta_{A_{i}} V_{A_{i}}}
\right)
:= \sigma_{\eta_{A_{i}}}^{2},
\end{equation}

In the asymptotic regime ($m \gg 1$), the estimator $\hat{\eta}_{A_{i}}$ can be accurately approximated by a Gaussian distribution due to the central limit theorem. Therefore, a conservative lower bound on the transmissivity, compatible with a parameter estimation failure probability $\epsilon_{\mathrm{PE}}$, is given by

\begin{equation}\label{Eq:PE5}
\eta_{A_{i},\mathrm{min}}
\simeq
\eta_{A_{i}}
-
w \, \sigma_{\eta_{A_{i}}},
\end{equation}
where the confidence parameter $w$ is defined as

\begin{equation} \label{Eq:PE6}
w =\sqrt{2 \ln \left( \frac{1}{\epsilon_{\mathrm{PE}}} \right)}.
\end{equation}

Finally, extending the above result to the multiuser scenario, the worst-case estimator of the channel transmissivity associated with the $i^{\mathrm{th}}$ user is given by
\begin{equation}\label{Eq:PE7}
\eta_{A_{i},\mathrm{min}}
\simeq
\eta_{A_{i}}
-
\frac{2 w \eta_{A_{i}}}{\sqrt{m}}
\sqrt{
2 +
\frac{\sigma_{A_{i}}^{2}}
{\eta_{A_{i}} V_{A_{i}}}
}.
\end{equation}

In the considered multiuser configuration, all users access the same physical quantum channel. Consequently, the transmissivity is identical for each user, i.e., $\eta_{A_{i}}=\eta$, which corresponds to the transmissivity of the common quantum channel. Under this assumption, the worst-case estimator of the overall channel transmissivity reduces to

\begin{equation}\label{Eq:PE8}
\eta_{\mathrm{min}}
\simeq
\eta
-
\frac{2 w \eta}{\sqrt{m}}
\sqrt{
2 +
\frac{\sigma_{A_{i}}^{2}}
{\frac{\eta}{N^{2}} V_{A_{i}}}
}.
\end{equation}

\subsubsection{Estimation of the Noise Variance}

Using the disclosed data samples, we estimate the variance of the noise variable $z_{A_{i}}$ by constructing the maximum likelihood estimator $\hat{\sigma}^{2}_{A_{i}}$, defined as \cite{PhysRevResearch.6.023321}

\begin{equation}\label{Eq:PE8}
\hat{\sigma}^{2}_{A_{i}}
=
\frac{1}{m}
\sum_{l=1}^{m}
\left(
X_{B_{i}l}
-
\sqrt{\hat{\eta}_{A_{i}}}\, X_{A_{i}l}
\right)^{2}.
\end{equation}
Under the assumption of Gaussian noise, the estimator $\hat{\sigma}^{2}_{A_{i}}$ follows a scaled chi-squared distribution with mean and variance given by

\begin{equation}\label{Eq:PE9}
\mathbb{E}[\hat{\sigma}^{2}_{A_{i}}]
=
\sigma^{2}_{A_{i}},
\qquad
\mathrm{Var}(\hat{\sigma}^{2}_{A_{i}})
=
\frac{2\sigma_{A_{i}}^{4}}{m}.
\end{equation}

For a sufficiently large number of samples ($m \gg 1$), $\hat{\sigma}^{2}_{A_{i}}$ can be approximated by a Gaussian distribution. Therefore, a conservative upper bound on the noise variance, compatible with a parameter estimation failure probability $\epsilon_{\mathrm{PE}}$, is given by

\begin{equation}\label{Eq:PE10}
\sigma^{2}_{A_{i}, \mathrm{max}}
\simeq
\sigma^{2}_{A_{i}}
+
w \, \sigma^{2}_{A_{i}}
\sqrt{\frac{2}{m}},
\end{equation}
where 
\begin{align}
    \sigma^{2}_{A_{i}} &=  \frac{\eta}{N^2} \sum_{\substack{k=1 \\ k \neq i}}^{N} M_{i} M_{k}\, V_{A_{k}}
+ \frac{(1-\eta)}{N} M_{i} W \nonumber\\
&\quad + M_{i} \left( \sum_{r=1}^{q} \frac{(-1)^{p_r(i)}}{\sqrt{2^{\,q-r+1}}} \right)^{2}\sigma
\end{align}
Finally, one can compute a new covariance matrix for $\text{Eve}$ with 
the worst-case parameter estimators, which should be used to compute the expected secret key rate in the finite-size regime. For the Holevo bound, it can be simplified as
\begin{equation}\label{Eq:61}
 \chi_{\epsilon_{\mathrm{PE}}}(X_{B_{i}}:E) = \sum^{2}_{n=1}g(\tilde{\nu}_{n}) - \sum^{4}_{n=3}g(\tilde{\nu}_{n}) 
 \end{equation}
where $\tilde{\nu}_{n}$ are the new symplectic eigenvalues of the corresponding covariance matrices given by the homodyne detection after the parameter estimation. The symplectic eigenvalues evaluated using the worst-case parameter estimators are given by

\begin{align}
\tilde{\nu}_{1,2}
&= \frac{1}{2}
\left[
\sqrt{(E^{\mathrm{est}}_{V} + W)^{2} - 4\eta_{\mathrm{min}}(W^{2} -1)}
\right. \nonumber\\
&\quad \left.
\pm (E^{\mathrm{est}}_{V} - W)
\right], \label{Eq:62} \\
\tilde{\nu}_{3,4}
&= \sqrt{
\frac{1}{2}
\left(
\Delta^{\mathrm{est}} \pm \sqrt{(\Delta^{\mathrm{est}})^{2} - 4\mu^{\mathrm{est}}}
\right)
}. \label{Eq:63}
\end{align}
where
\begin{align}
\Delta^{\mathrm{est}}&=W^{2}+\left( E_{V}^{\mathrm{est}} \right)^{2}+2\, \Psi^{\mathrm{est}} \Gamma^{\mathrm{est}}_{i}
-\frac{ W \left( \Psi^{\mathrm{est}} \right)^{2} }
{ \beta_{i}^{\mathrm{est}} }\nonumber\\
&\quad-\frac{E_{V}^{\mathrm{est}}
\left( \Gamma^{\mathrm{est}}_{i} \right)^{2}}
{ \beta_{i}^{\mathrm{est}} }-2 \left(\Phi^{\mathrm{est}} \right)^{2},\\
\mu^{\mathrm{est}}&=\frac{W E_{V}^{\mathrm{est}}-
\left( \Phi^{\mathrm{est}} \right)^{2}}{\beta_{i}^{\mathrm{est}}}\nonumber\\
&\quad \times\Bigg(W E_{V}^{\mathrm{est}} \beta_{i}^{\mathrm{est}}-W \left( \Gamma^{\mathrm{est}}_{i} \right)^{2}-E_{V}^{\mathrm{est}} \left( \Psi^{\mathrm{est}} \right)^{2}\nonumber\\
&\qquad-\left( \Phi^{\mathrm{est}} \right)^{2}
\beta_{i}^{\mathrm{est}}+2\, \Phi^{\mathrm{est}}
\Gamma^{\mathrm{est}}_{i}\Psi^{\mathrm{est}}\Bigg).
\end{align}

Additionally, we define:
\begin{align}
E_{V}^{\mathrm{est}}&=\left( 1 - \eta_{\mathrm{min}} \right) V_{T}+\eta\, W\\
\Gamma^{\mathrm{est}}_{i} &= \sqrt{\frac{\eta_{\mathrm{min}}(1-\eta_{\mathrm{min}})}{N} M_{i}} W 
- \sqrt{\frac{\eta_{\mathrm{min}}(1-\eta_{\mathrm{min}})M_{i}}{N^{3}}} V_{A_{i}} \nonumber \\ &\quad -\sqrt{\frac{\eta_{\mathrm{min}}(1-\eta_{\mathrm{min}})M_{i}}{N^{3}}} \sum^{N}_{k\neq i} M_{k} V_{A_{k}}, \\
\Psi^{\mathrm{est}}
&= \sqrt{\frac{(1 - \eta_{\mathrm{min}})M_{i}}{N}}\; \sqrt{W^{2} - 1}\\
 \Phi^{\mathrm{est}} &= \sqrt{\eta_{\mathrm{min}} (W^{2} - 1)}\\
\beta^{\mathrm{est}}_{i} &= \frac{\eta_{\mathrm{min}}}{N^2} V_{A_{i}} + \sigma^{2}_{A_{i}, \mathrm{max}}
\end{align}

\section{Results and Numerical Analysis}\label{Sec:5}

In this section, we present comprehensive numerical simulation results of the SKR of the proposed q-CDMA-based CV-QKD system. The impact of important system parameters—including the correction factor $M$, reconciliation efficiency $\beta$, inter-user interference noise, and environmental noise—on both the asymptotic and finite-size SKRs is systematically investigated. These results provide quantitative insights into the fundamental factors governing the performance and scalability of the q-CDMA-based CV-QKD architecture.

We assume that each transmitter, denoted by $\text{Alice}_{i}$, employs an identical signal power to encode the raw key information. According to the input–output relations of the annihilation operators given in Eq.~(\ref{Eq:13C}), the interference experienced by the $i^{\mathrm{th}}$ user arises from the modes $\hat{a}_{A_k}$ with $k \neq i$, corresponding to the input modes of all other users $\text{Alice}_{k}$. Consequently, the variance of the interference modes is assumed to be identical to that of the signal mode, i.e., $V_{A_k} = V_{A_i}$.

Throughout the simulations, the variance of the preparation noise modes is fixed to $V_0 = 1$, measured in shot-noise units $(\mathrm{SNU})$, for both $\text{Alice}_{i}$ and $\text{Alice}_{k}$. Furthermore, the variance of Eve’s quadrature is set to $W = 1$, corresponding to a pure-loss channel model. This assumption effectively isolates the contribution of trusted preparation noise and represents the optimal collective Gaussian attack, thereby establishing a conservative yet physically realistic security benchmark. Security proven under the condition $W = 1$ is therefore expected to remain qualitatively valid in the presence of small excess channel noise \cite{PhysRevLett.105.110501}. In addition, the environmental noise variance is fixed at $\sigma = 1~\mathrm{SNU}$, accounting solely for the vacuum fluctuations entering through the unused port of the beam splitter. The attenuation coefficient for standard optical fiber is taken as $\alpha = 0.25~\mathrm{dB/km}$, and the reconciliation efficiency is set to $\beta = 0.95$, consistent with experimentally reported values \cite{Liu:24}. Moreover, we assume that all chaotic phase shifters $\text{CPS}_{i}$ are perfectly synchronized with their corresponding counterparts $\text{CPS}^{'}_{i}$, ensuring identical correction factors for all users, i.e., $M_i = M_k = M$.

\begin{figure}[htp]
   \centering
\includegraphics[width=0.5\textwidth]{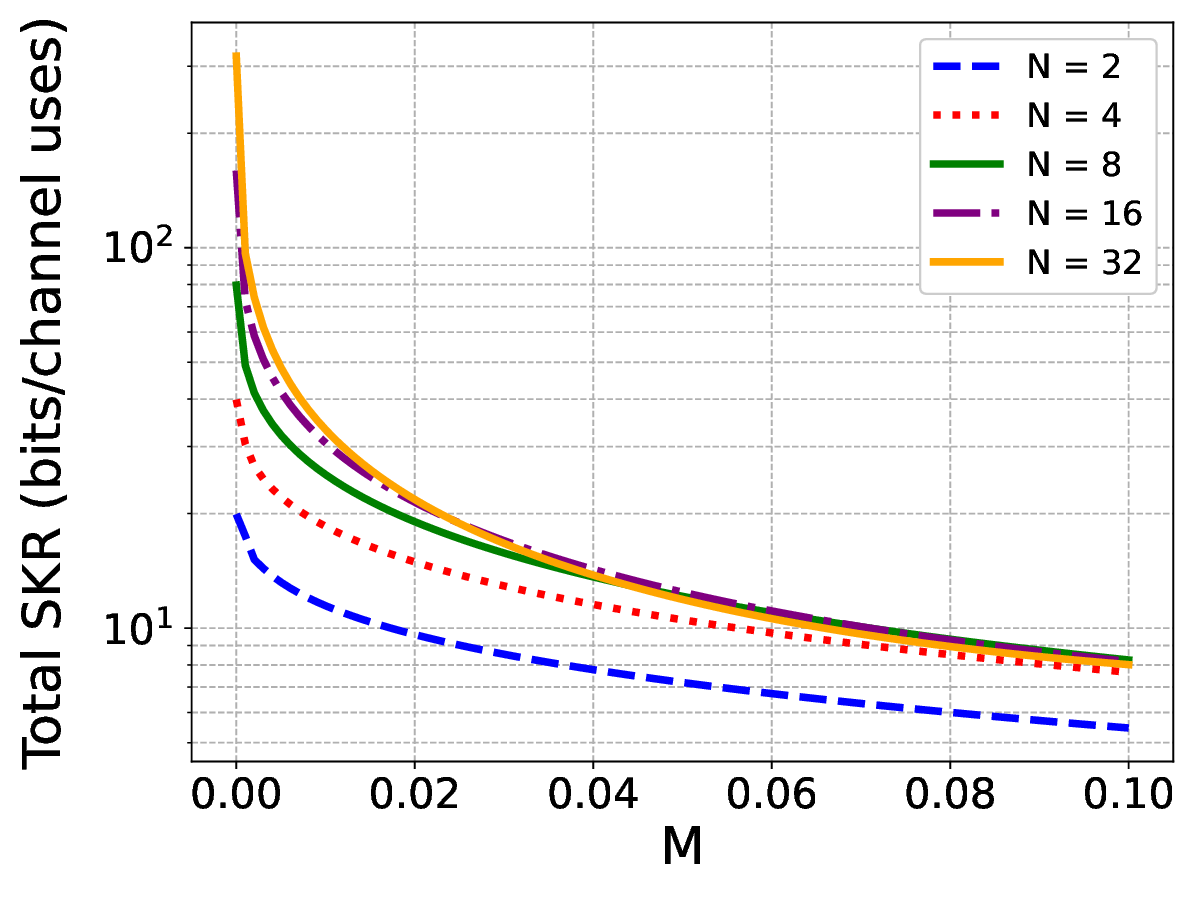}
         \caption{ Total asymptotic SKR as a function of correction factor $M$ for the q-CDMA CV-QKD system. Other simulation parameters are, $V_{0} = 1$ $\mathrm{SNU}$, $V_{S_{i}} = 10^{3}$ $\mathrm{SNU}$, $d= 100 \, \text{Km}$, $\alpha = 0.25 \, \text{dB/Km}$, $W =1$ $\mathrm{SNU}$ and $\sigma = 1$ $\mathrm{SNU}$.}
     \label{Fig:skrm}
\end{figure}

\begin{figure}[htp]
   \centering
\includegraphics[width=0.45\textwidth]{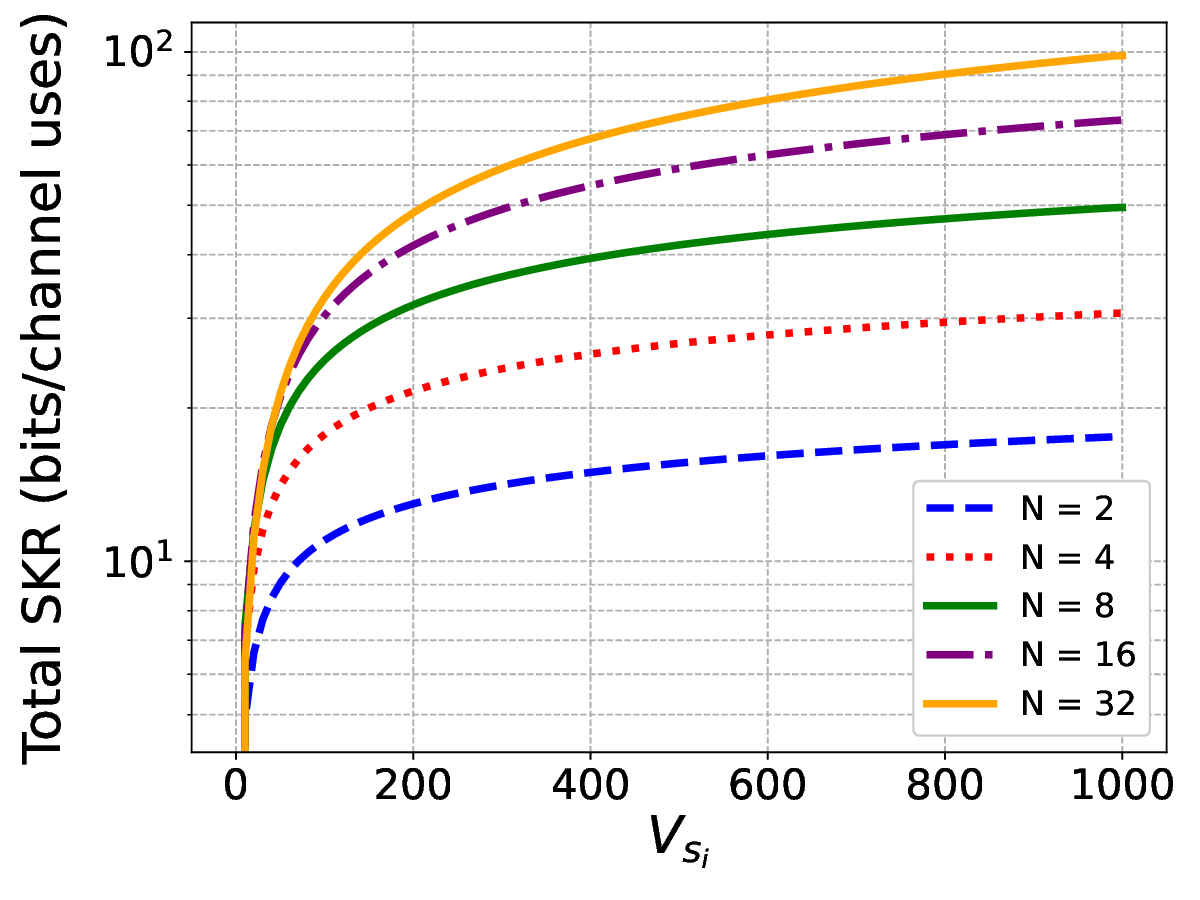}
         \caption{Total asymptotic SKR as a function of modulation variance $V_{S_{i}}$ of $\text{Alice}_{i}$ for the q-CDMA CV-QKD system at, $d = 100\, \text{Km}$. Other simulation parameters are the same as those of Fig.\ \ref{Fig:skrm}. } 
     \label{fig:skrv}
\end{figure}

\begin{figure*}[htp]
\centering
\begin{subfigure}{0.45\textwidth}
    \centering
\includegraphics[width=\textwidth]{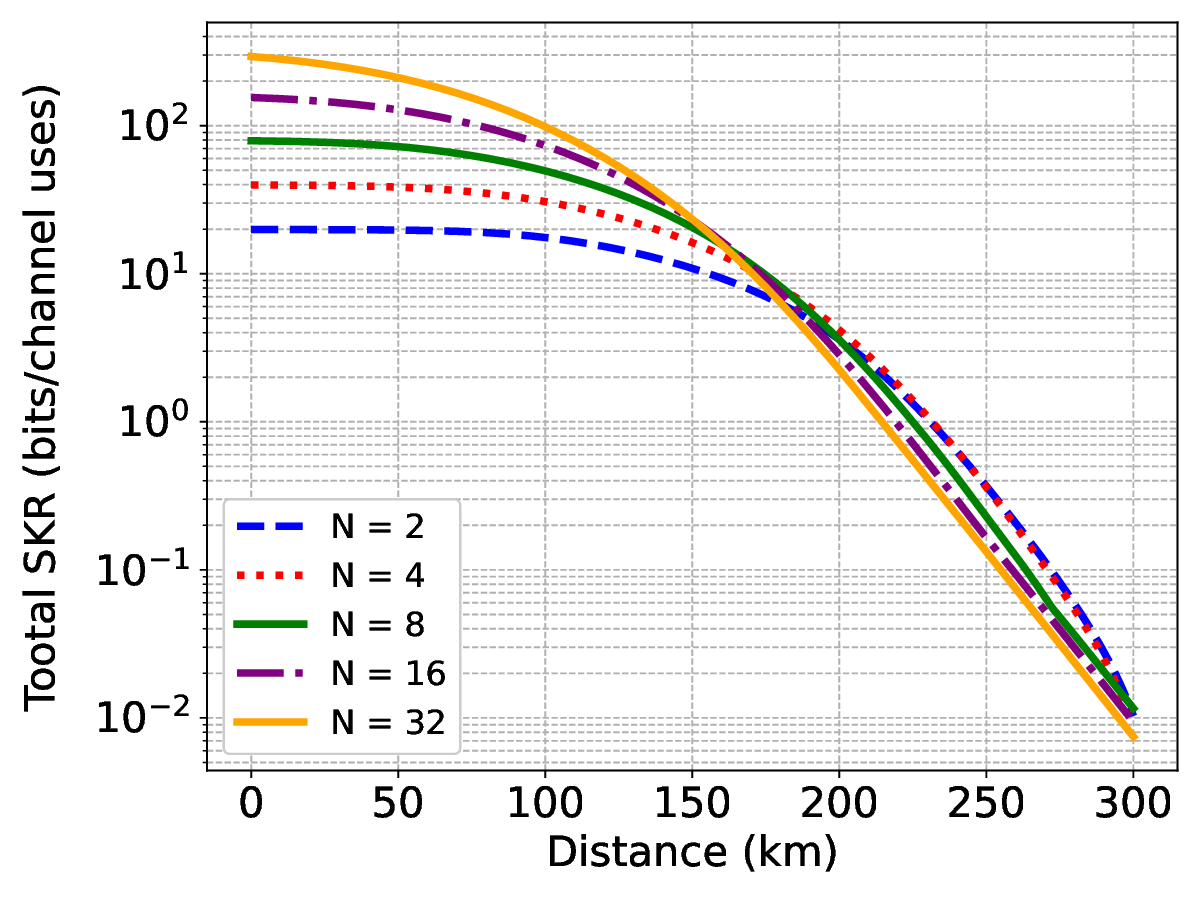}
    \caption{Total secret key rate}
    \label{fig6_a}
\end{subfigure}
\begin{subfigure}{0.45\textwidth}
    \centering
\includegraphics[width=\textwidth]{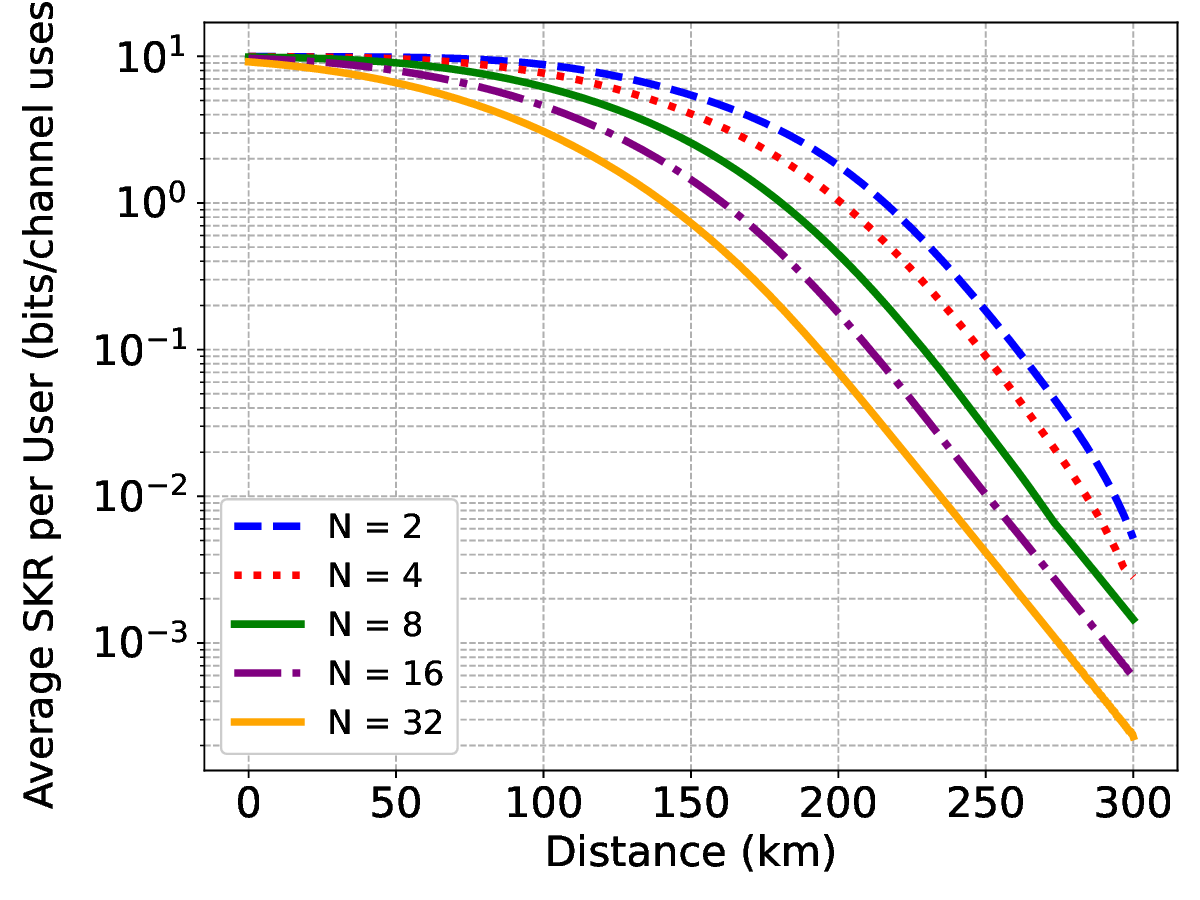}
    \caption{Average secret key rate per user}
    \label{fig:N1}
\end{subfigure}
\caption{ Asymptotic total SKR and average SKR per user as a function of transmission distance $d$ between $\text{Alice}_{i}$ and $\text{Bob}_{i}$, for different user pairs $N$ with fixed signal variance $V_{S_{i}}$. Other simulation parameters are the same as those of Fig.\ \ref{Fig:skrm}.}
\label{fig6}
\end{figure*}

\begin{figure*}[htp]
\centering
\begin{subfigure}{0.45\textwidth}
    \centering
    \includegraphics[width=\textwidth]{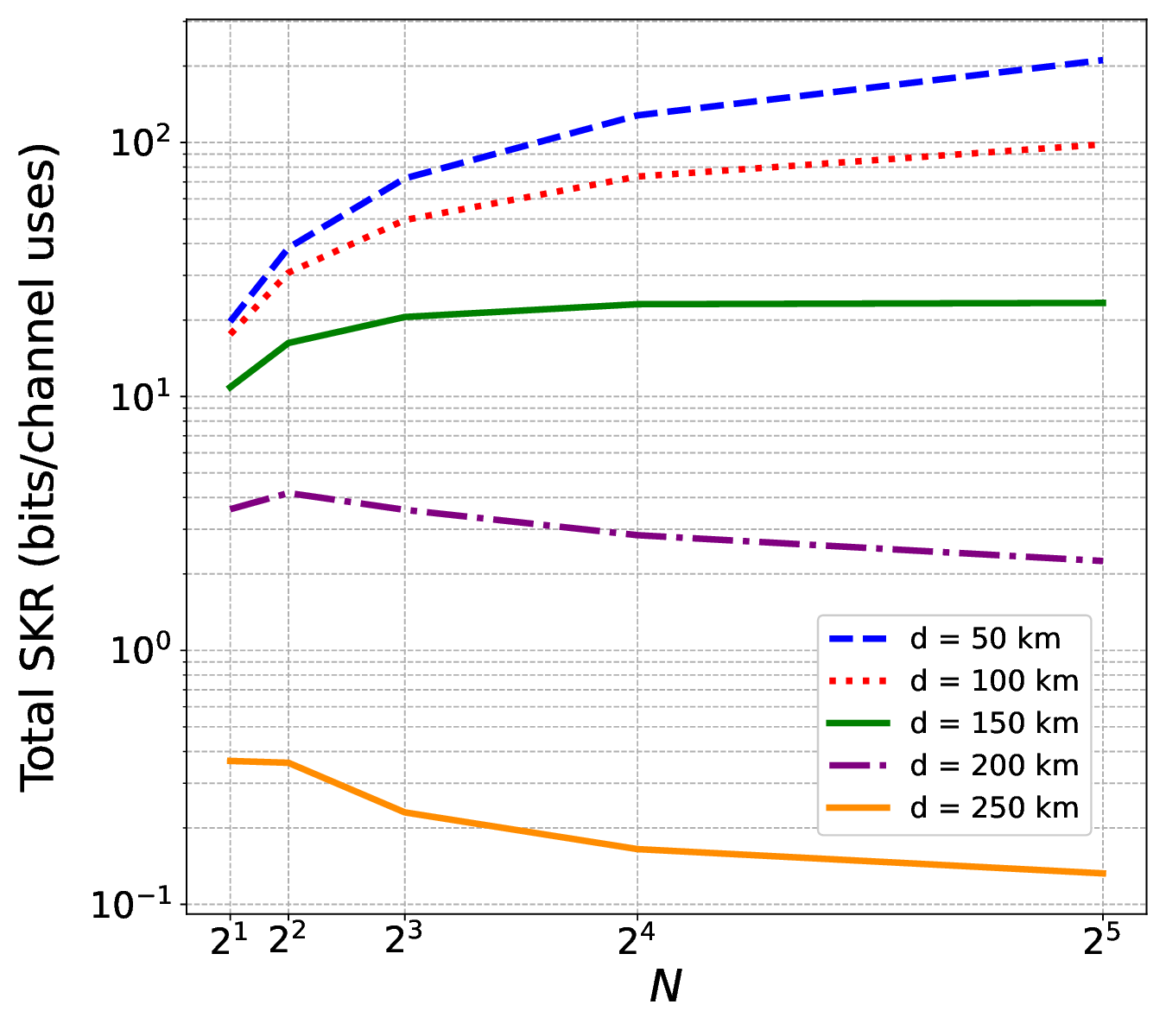}
    \caption{Total secret key rate}
    \label{fig:N2a}
\end{subfigure}
\begin{subfigure}{0.45\textwidth}
    \centering
    \includegraphics[width=\textwidth]{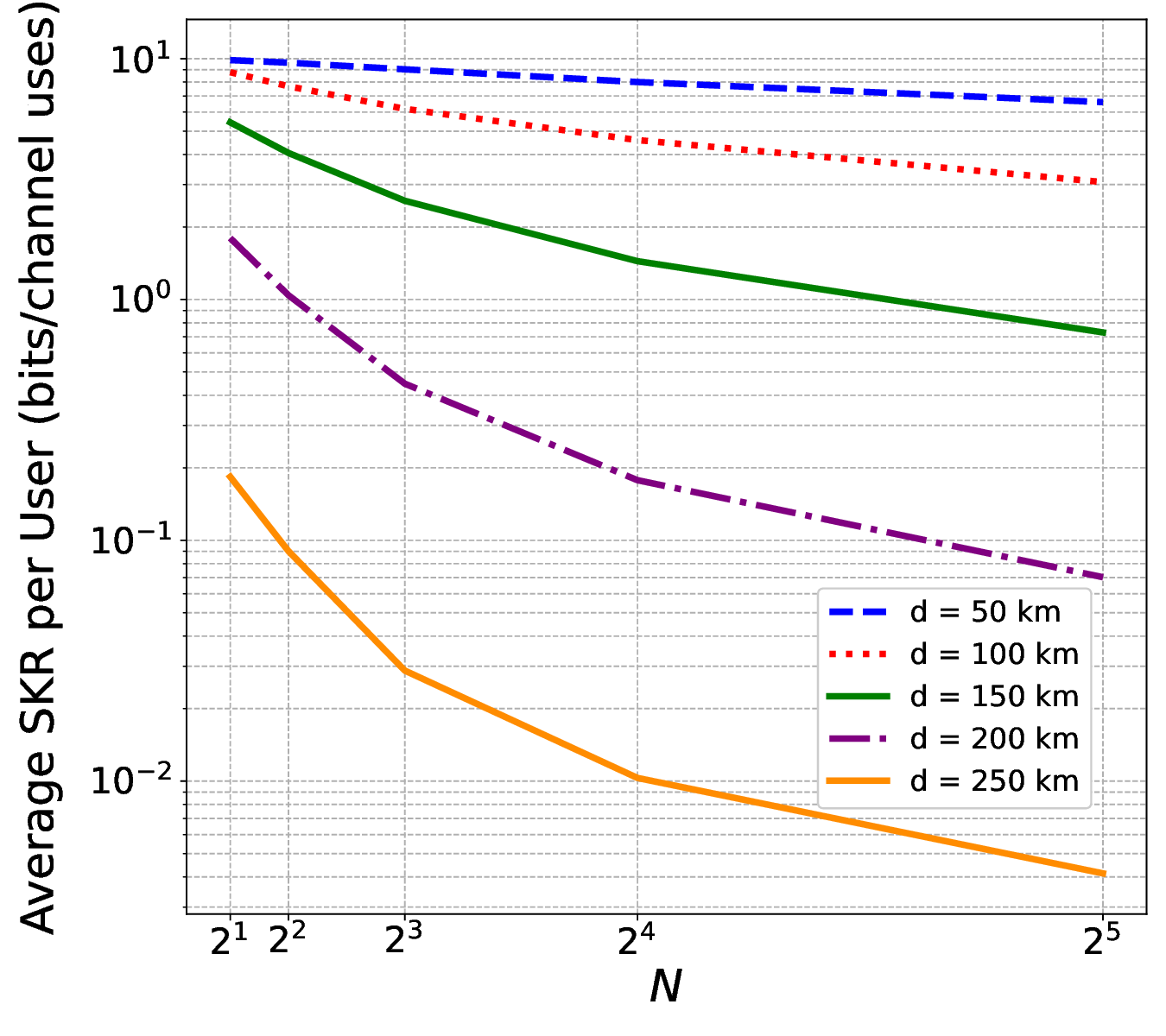}
    \caption{Average secret key rate per user}
    \label{fig:N2b}
\end{subfigure}
\caption{ Asymptotic total SKR and average SKR per user as a function of the number of users $N$ for different transmission distances and fixed $V_{S_{i}}=1000$ $\mathrm{SNU}$. Other simulation parameters are the same as those of Fig.\ \ref{Fig:skrm}.}
 \label{fig:N2}
\end{figure*}
\begin{figure*}[htp]
\centering
\begin{subfigure}{0.45\textwidth}\centering
 \includegraphics[width=\textwidth]{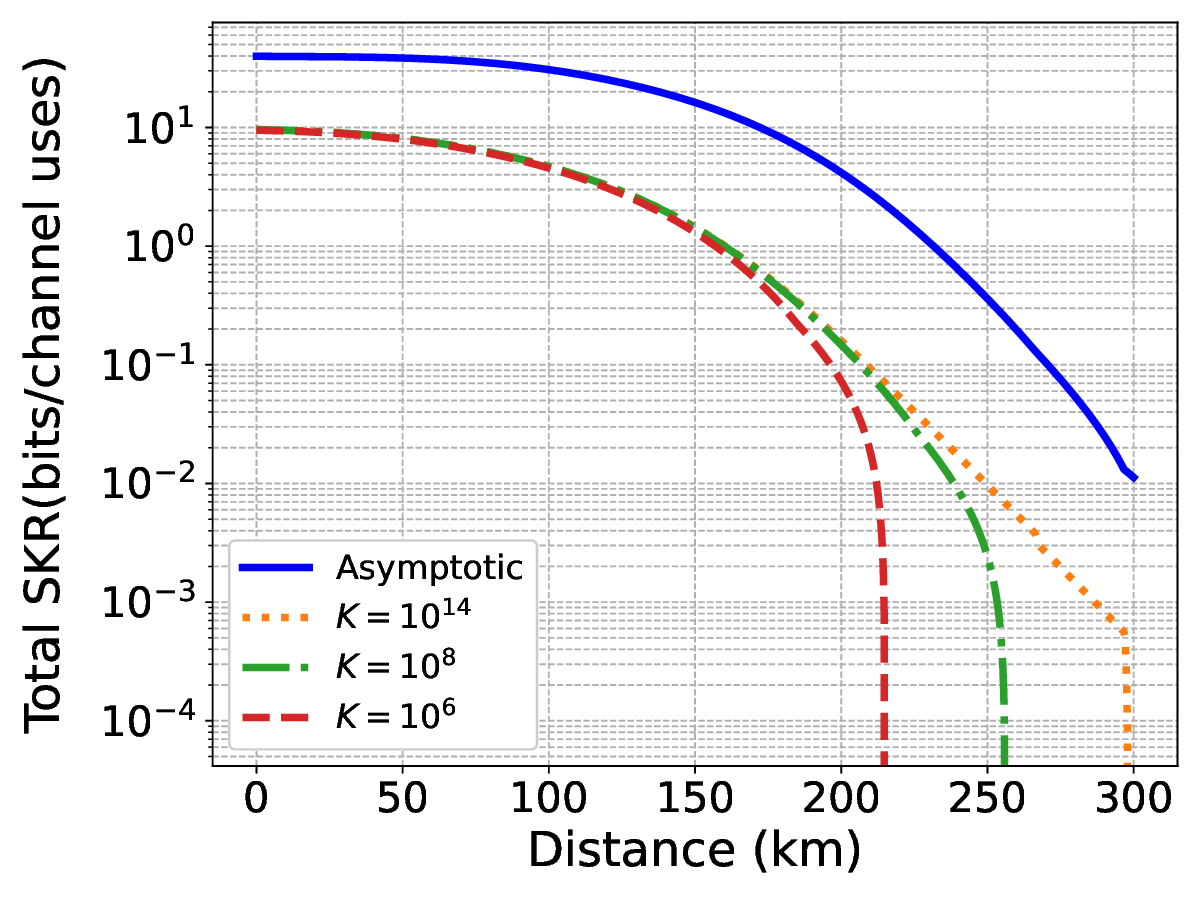}
 \caption{$N = 4$ User Pairs}
\end{subfigure}
\begin{subfigure}{0.45\textwidth}\centering
  \includegraphics[width=\textwidth]{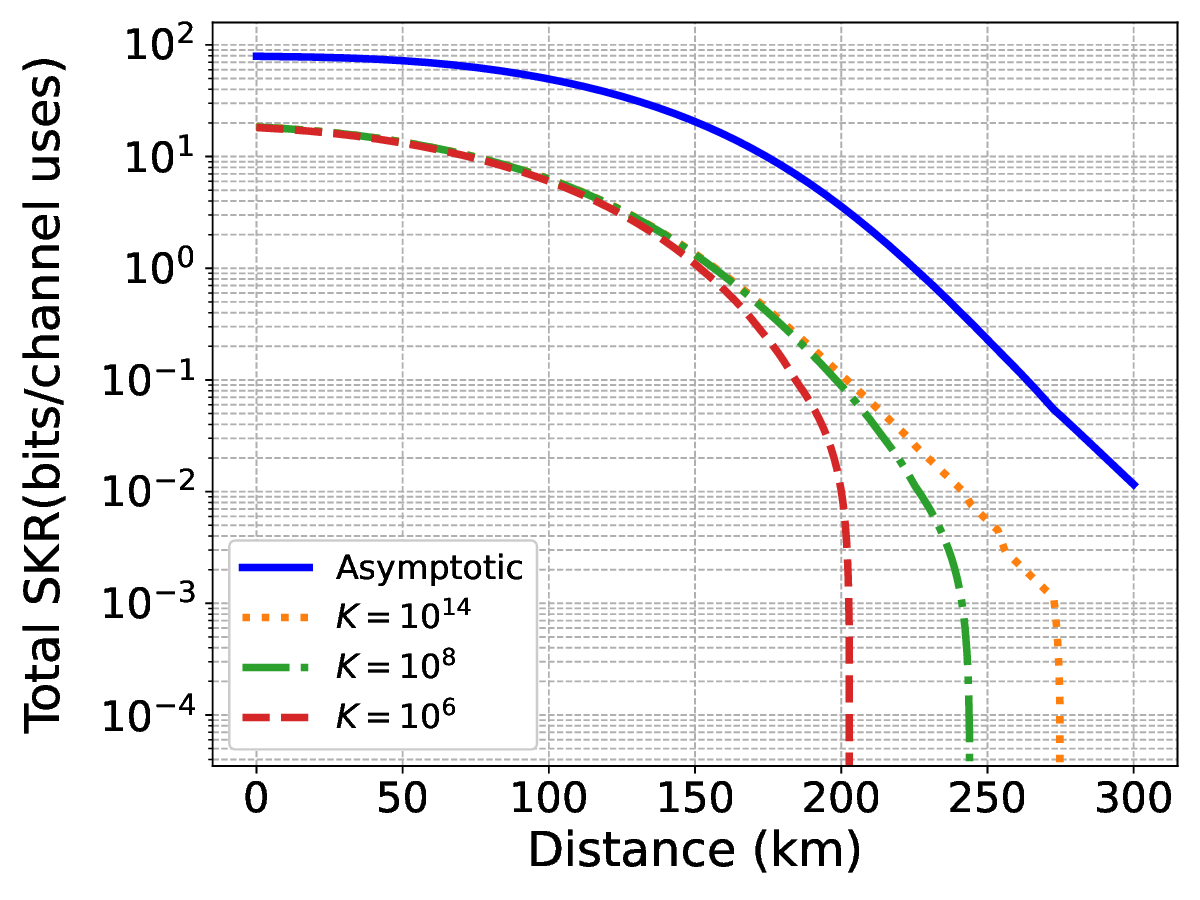}
   \caption{$N = 8$ User Pairs}
 \end{subfigure}\\
\begin{subfigure}{0.45\textwidth}\centering
 \includegraphics[width=\textwidth]{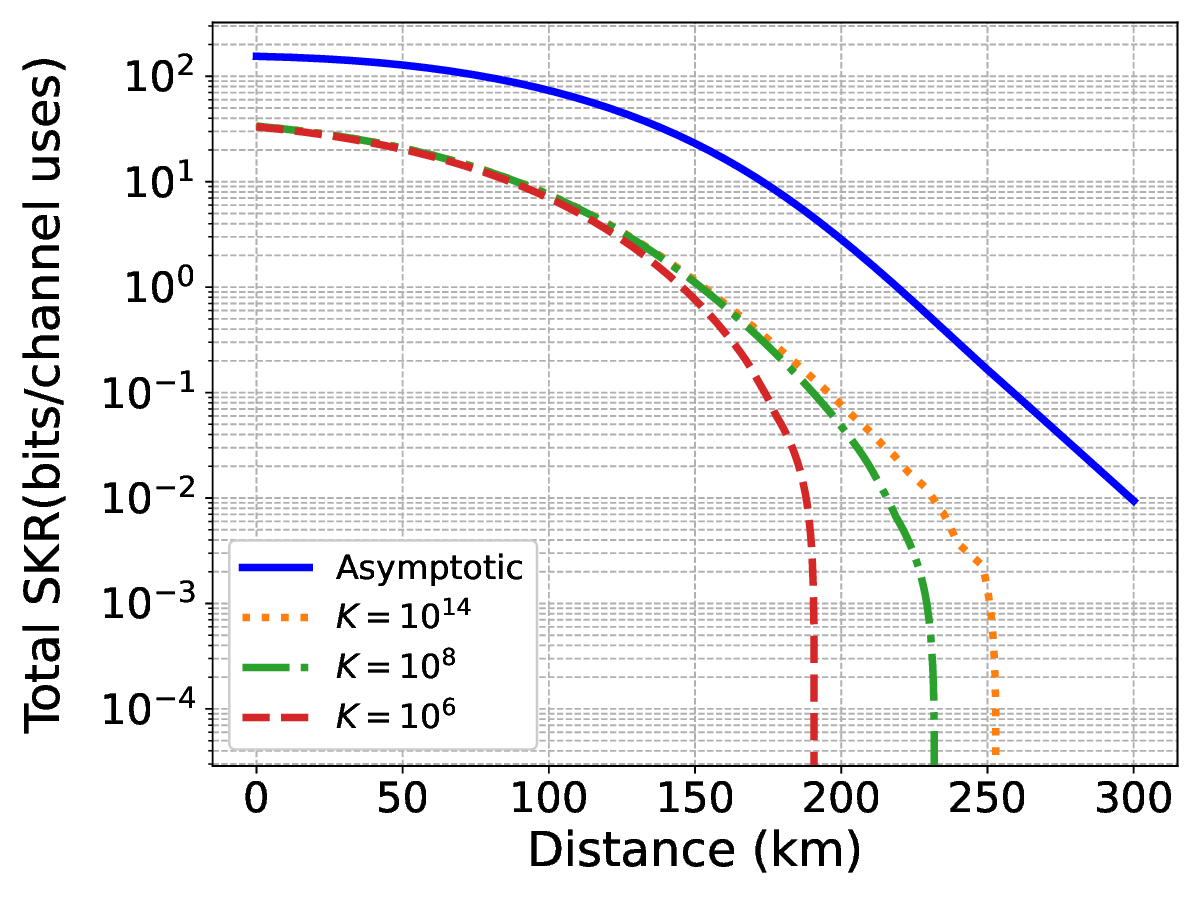}
   \caption{$N = 16$ User Pairs}
 \end{subfigure}
\begin{subfigure}{0.45\textwidth}\centering
 \includegraphics[width=\textwidth]{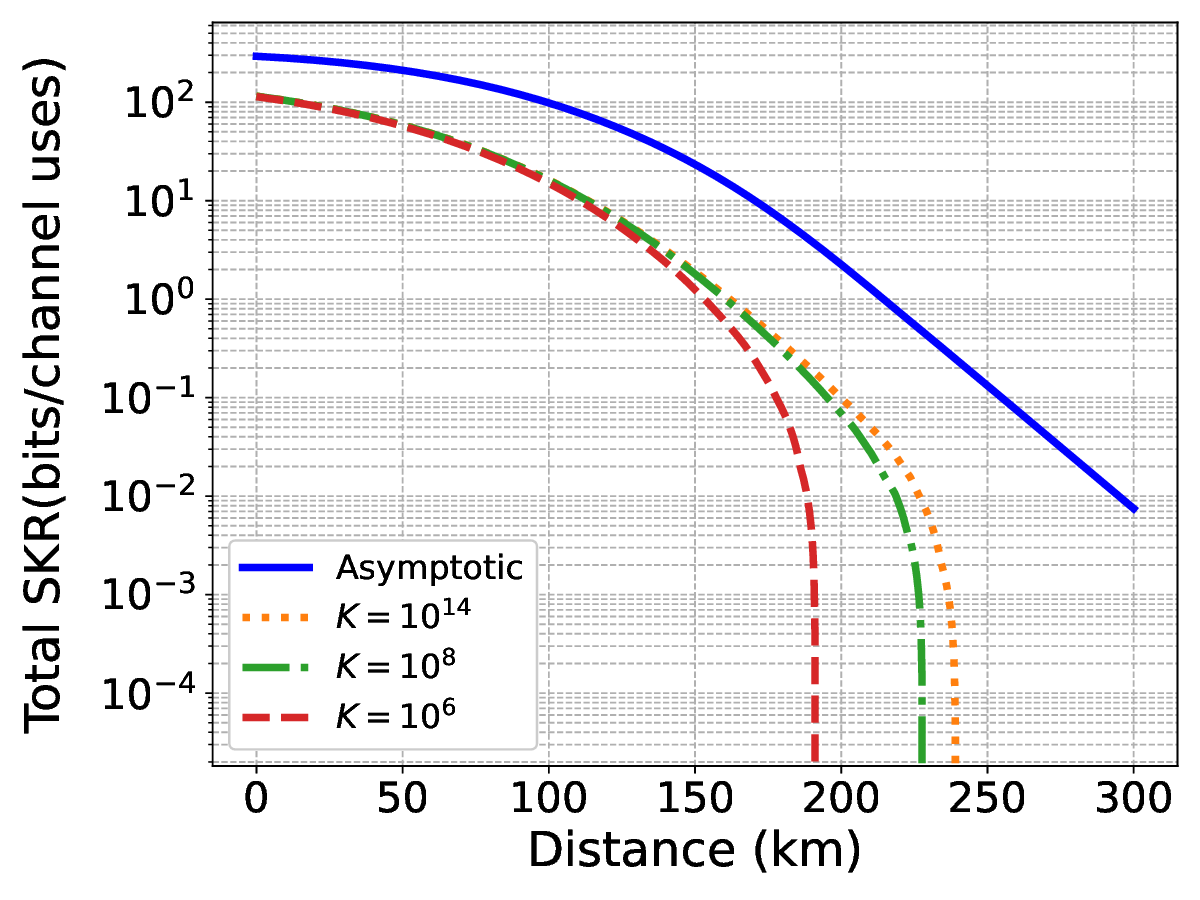}
  \caption{$N = 32$ User Pairs}
\end{subfigure}\hfill
\caption{Total SKR with finite size effect as a function of transmission distances between $\text{Alice}_{i}$ and $\text{Bob}_{i}$ for the block length $ K = 10^{6},  10^{8}, 10^{14}$  for the different user pairs and $\beta = 0.98$ Other simulation parameters are the same as those of Fig.\ \ref{Fig:skrm}.}
\label{fig:N3}
\end{figure*}

Fig.~\ref{Fig:skrm} shows the total asymptotic SKR, expressed in bits per channel use, as a function of the correction factor \(M\) for different numbers of user pairs \(N\). The parameter \(M\) quantifies the degree of chaotic dynamics in the q-CDMA system and is uniquely determined by the power spectral density of the chaotic modulation, together with the lower and upper bounds of its frequency band. Previous studies have shown that broadband chaotic spectra typically lead to smaller values of \(M\) \cite{PhysRevA.92.042327}. It is evident that, for each user pair, the SKR attains significantly larger values for smaller correction factors \(M\). This behavior can be directly understood from the input--output quadrature relations given in Eq.~(\ref{Eq:13C}). In particular, a reduction in \(M\) suppresses multiuser interference noise as well as the contribution from environmental noise, thereby enhancing the effective correlations between the quadratures of \(\text{Alice}_{i}\) and \(\text{Bob}_{i}\). As discussed in Sec.~\ref{Sec:2}, a smaller \(M\) implies that the quadrature components of \(\text{Alice}_{i}\) more closely match those of \(\text{Bob}_{i}\), which mitigates cross-talk effects and leads to an improved SKR. Moreover, for a fixed value of \(M\), the total SKR increases with the number of user pairs \(N\). Furthermore, it can be observed that for $N>2$ users the total SKR saturates as \(M\) increases. This saturation behavior arises because, in the multiuser regime, interference noise and environmental noise dominate the system performance, thereby limiting further improvements in the SKR.

Fig.~\ref{fig:skrv} illustrates the total SKR as a function of the transmit signal variance $V_{S_i}$ for different user pairs $N$, at a fixed transmission distance of $d = 100~\mathrm{km}$. The results indicate that the total SKR initially increases with increasing $V_{S_i}$ and then saturates for large $V_{S_i}$. This observation is also analytically supported by (\ref{Eq:20a}), where the variance $V_{S_i}$ appears both in the first term corresponding to the signal component and the summation term accounting for the interference contributions from all $k\neq i$ users. This behavior suggests that aggregate interference from all active users significantly impacts the achievable SKR.

Fig.~\ref{fig6} illustrates the total asymptotic SKR and the average SKR per user versus the transmission distance $d$ (in $\text{km}$) between $\text{Alice}_{i}$ and $\text{Bob}_{i}$, for different user pair $N$ with fixed signal variance $V_{S_{i}}=10^{3}~\mathrm{SNU}$. It can be observed from Fig. ~\ref{fig6_a} that in a q-CDMA network, the total SKR decreases with distance for all user pairs $N$. Furthermore, we observe that for smaller distances, the total SKR is larger for a higher number of user pairs $N$; however, beyond a threshold distance, the total SKR decreases for a larger number of user pairs $N$. This decrease in total SKR occurs because, for larger transmission distances, the signal component decreases and the interference noise dominates for a higher number of user pairs $N$. Additionally, from Fig.~\ref{fig:N1}, it can be observed that the average SKR per user decreases as the number of user pairs $N$ increases for all transmission distances.

Fig.~\ref{fig:N2} shows the asymptotic total SKR and the average SKR per user as a function of the number of user pairs, $N$, for different transmission distances. From Fig.~\ref{fig:N2a}, it can be observed that the total SKR initially increases as the number of user pairs $N$ increase; however, beyond a threshold $N$, the total SKR starts decreasing, with the decrease more pronounced for larger transmission distances.  This trend can be explained using Eq.~(\ref{Eq:16}), where the total SKR is the sum of the individual SKRs of all user pairs. According to Eq.~(\ref{Eq:13C}), the input signal of each intended $\text{Alice}_{i}$ is attenuated by a factor of $N^{2}$, while the associated interference noise terms are reduced by a factor proportional to $N$ for all users. As a result, the total SKR initially benefits from the increasing number of active user pairs, as reflected in Eq.~(\ref{Eq:16}). However, as $N$ continues to grow, the attenuation of individual user signal contributions becomes dominant, leading to saturation and eventual degradation of the total SKR. Furthermore, Fig.~\ref{fig:N2b} shows that the average SKR per user decreases more rapidly with increasing $N$ at larger transmission distances, indicating the stronger impact of channel loss and reduced signal strength in long-distance quantum communication.

Finally, the impact of finite-size effects on the SKR has been numerically studied by comparing the asymptotic SKR with the finite-size SKR as a function of the transmission distance. In the finite-size regime, block sizes of $K = 10^{6}$, $10^{8}$, and $10^{14}$ are considered, while the asymptotic case assumes $K \to \infty$, the security threshold parameters are fixed as $\bar{\epsilon}=\epsilon_\mathrm{PA}= 10^{-10}$ and the number of measurements used for parameter estimation are $n=m=\frac{K}{2}$, to ensure a stringent balance between security and reliability \cite{PhysRevA.81.062343}. As illustrated in Fig.~\ref{fig:N3}, both asymptotic and finite-size SKRs exhibit a monotonic decrease with increasing transmission distance. Moreover, the finite-size SKR decreases more rapidly than the asymptotic SKR with increasing transmission distance between the user pairs. This declination is attributed to increased losses and reduced quantum signal strength over long distances. This effect is further amplified as the number of user pairs increases, demonstrating that multiuser interference, combined with finite-size constraints, significantly limits the SKR performance in long-distance scenarios.

\section{Conclusion}\label{Sec:6}


In summary, we have proposed a novel q‑CDMA framework for CV‑QKD networks. The proposed scheme enables multiple transmitters to simultaneously share a common quantum channel by employing chaotic phase shifters for quantum‑state encoding and decoding. A binary‑tree‑structured beam‑splitter network is introduced to realize efficient multiplexing and demultiplexing of quantum signals among $N$ user pairs. The quadrature input–output relations for all sender–receiver pairs are derived, and the corresponding SKR performance is analyzed by accounting for the effects of chaotic modulation and synchronization, multiuser interference, and environmental noise. Extensive numerical simulations quantify the influence of critical system parameters, including the correction factor $M$, reconciliation efficiency $\beta$, multiuser interference noise, and environmental noise, on the achievable SKR. The results further demonstrate that the finite‑size SKR degrades more rapidly with increasing transmission distance than its asymptotic counterpart, due to enhanced channel loss and reduced quantum signal strength, which adversely affect parameter estimation accuracy. Overall, the proposed q‑CDMA framework is established as a promising architecture for high‑rate CV‑QKD in multiuser environments, thereby paving the way for realizing scalable and secure long-distance (globally-accessible) quantum communication networks.

\section{Acknowledgements}\label{Sec:7}

The authors thank Cosmo Lupo and Nitin Jain for insightful discussions.


\bibliographystyle{IEEEtran}
\bibliography{qcdma}

\end{document}